\documentclass[pre,twocolumn]{revtex4-2}
\usepackage{graphicx}
\graphicspath{{images/}}
\usepackage{algorithm}
\usepackage{algpseudocode}
\usepackage{amsmath}
\usepackage{amssymb}
\usepackage{xcolor}

\newcommand{\beq}{\begin{equation}}
\newcommand{\eeq}{\end{equation}}

\newcommand{\ba}{\begin{array}}
\newcommand{\ea}{\end{array}}

\newcommand{\bea}{\begin{eqnarray}}
\newcommand{\eea}{\end{eqnarray}}

\newcommand{\bc}{\begin{center}}
\newcommand{\ec}{\end{center}}

\newcommand{\bt}{\begin{tabular}}
\newcommand{\et}{\end{tabular}}

\newcommand{\bi}{\begin{itemize}}
\newcommand{\ei}{\end{itemize}}

\newcommand{\bd}{\begin{description}}
\newcommand{\ed}{\end{description}}

\newcommand{\bp}{\begin{pmatrix}}
\newcommand{\ep}{\end{pmatrix}}

\begin{document}
\title{Detection of Functional Communities in Networks of Randomly Coupled Oscillators Using the Dynamic-Mode Decomposition}

\author{Christopher W. Curtis}
\affiliation{Department of Mathematics and Statistics, San Diego State University}
\email[]{ccurtis@sdsu.edu}
\author{Mason A. Porter}
\affiliation{Department of Mathematics, University of California, Los Angeles}
\affiliation{Santa Fe Institute}

\begin{abstract}

Dynamic-mode decomposition (DMD) is a versatile framework for model-free analysis of time series that are generated by dynamical systems. We develop a DMD-based algorithm to investigate the formation of ``functional communities'' in networks of coupled, heterogeneous Kuramoto oscillators. In these functional communities, the oscillators in the network have similar dynamics. We consider two common random-graph models (Watts--Strogatz networks and Barab\'asi--Albert networks) with different amounts of heterogeneities among the oscillators. In our computations, we find that membership in a community reflects the extent to which there is establishment and sustainment of locking between oscillators.  We construct forest graphs that illustrate the complex ways in which the heterogeneous oscillators associate and disassociate with each other.       
\end{abstract}

\maketitle



\section{Introduction} \label{sec1}

Researchers in myriad disciplines employ networks to represent entities (i.e., nodes) that interact with each other through their connections (which are encoded by edges) \cite{newman-book2018}. Network architecture, in turn, affects the dynamics of systems that evolve on networks \cite{porter2016dynamical}. Network structure can profoundly impact the spread of diseases \cite{kiss} and information \cite{lehman2018}, the collective behavior of coupled oscillators \cite{arenas2,rodrigues}, and more.

The analysis of networks plays an increasingly important role throughout the sciences and engineering, and this is very prominent in the study of dynamical systems \cite{newman-book2018,porter2}. A key question is how interacting entities in a network form collective structures like \emph{communities}, which can take the form of either {\it structural} or {\it functional} communities.  A \emph{structural community} is a dense set of nodes that is connected sparsely to other dense sets of nodes \cite{porter3,fortunato}. A very large number of approaches have been developed to algorithmically detect structural communities in networks. By contrast, considerably less effort has been devoted to the detection of \emph{functional communities}, which are based on the behaviors or dynamics of the nodes. One way to detect functional communities is by running a dynamical process on a network, constructing a new network (a so-called ``functional network") based on the time-series similarity of the outputs of the network's nodes, and then detecting structural communities in the functional network \cite{arenas1,bassett2013}. One can also detect functional communities using time-series output of experiments. Whether the time-series output comes from a model or experimental measurements, the focus in functional-community detection is the behavioral similarity of a network's entities over time.  These functional communities, whose name is inspired by studies of functional brain networks in neuroscience \cite{bassett2013}, arise from communities in a network in which the edges encode some type of time-series similarity between the nodes of the network. Another name for such communities is ``behavioral communities''. They were explored briefly in \cite{shalizi2007} using methods from information theory and structural community detection.

In the present study of functional communities, we consider the setting of coupled oscillators on networks. The network itself may have community structure (which is based on its structural communities), which one can examine through one or more of the myriad available methods for community detection \cite{fortunato}. Additionally, by examining the output of coupled oscillators on a network and tracking phenomena such as synchronization, one can also study the same system's functional community structure. This topic was explored by Arenas et al.~\cite{arenas1} in an investigation of coupled Kuramoto oscillators on networks. The fact that coupled Kuramoto models have been studied so extensively \cite{rodrigues} makes then an ideal test case for studies of functional-community detection in coupled oscillators, with a view towards extending such analysis to other systems (including experimental ones, such as in analysis of neuroimaging data \cite{bassett2013}). Although we study coupled oscillators in our work, other types of dynamical systems on networks can also have functional communities \cite{chauhan2012}.  

Arenas et al. \cite{arenas1} examined Kuramoto oscillators that are coupled to each other on a network with a hierarchical community structure (with smaller, denser communities nested inside larger, sparser ones), and they examined how the architecture of structural communities affects the formation of functional communities, as quantified by how long it takes the oscillators to synchronize.  They demonstrated that oscillators in denser communities synchronize faster than oscillators in sparser ones. Variations of coupled Kuramoto oscillators on networks have also been used to study the coalescence
of functional communities \cite{boccaletti,wu}. 
In these papers, the network architecture was fixed and the researchers sought to partition the networks based on the dynamics of the oscillators.

In our investigation of functional communities, we take a different perspective from those in the aforementioned papers.  We seek to detect communities from output dynamics, and we consider both the formation and the disappearance in time of functional communities.  To do this, we use the modal-decomposition technique that is known as the dynamic-mode decomposition (DMD).  For more information about DMD, see \cite{schmid,mezic1,kutz} for comprehensive reviews and \cite{taira, clainche, kutz7} for surveys and discussions of recent extensions. The primary benefit of the DMD is that it is a model-free data-processing tool that allows one to generate modal decompositions from arbitrarily complicated data sets over reasonably chosen time intervals.  
Moreover, in contrast to other modal-decomposition methods (such as principle-component analysis), DMD also gives a convenient mechanism to generate models from measured data alone, and it thereby gives a way to sidestep model development in situations in which it is difficult or even impossible.

In the present paper, instead of examining synchronization time scales as a way to partition a network of oscillators, we generate output data from coupled Kuramoto oscillators on networks (which we construct from random-graph models) and identify functional communities using modes that are generated by the DMD. The flexibility of the DMD allows us to do this in a time-dependent way, and it therey makes it possible to track the formation and disappearance of communities.


Our DMD-based approach provides a straightforward and flexible way to generate functional communities from a time series of the nodes of a network.  In our study of coupled oscillators, the formation of such communities still relies on synchronization of subsets of the oscillators, but our approach does not require the observation of any global attractors (which do not manifest on the time scales that we examine).  Therefore, we develop a flexible and adaptive method for identifying functional communities in large sets of time series of coupled nonlinear oscillators, and we anticipate that it will also be applicable to wider classes of time series in which community formation is of interest. Our work complements the recent results of Kunert-Graf et al. \cite{kutz5}, who used unsupervised learning techniques to cluster DMD modes to capture spatially and temporarily coherent patterns in complex signals. Our code is available at \url{https://github.com/cwcurtis/DMD_Community_Detect}.                

Our paper proceeds as follows. In Section \ref{sec2}, we give the necessary background and definitions to understand both DMD and how we generate our data. In Section \ref{sec3}, we explain our algorithm for the detection of functional communities. In Section \ref{sec4}, we present the results of our numerical experiments. Finally, in Section \ref{sec5}, we summarize our results and and discuss future work.

\section{Background and Definitions} \label{sec2}

\subsection{Dynamic-Mode Decomposition (DMD)} \label{dmd}

Consider a nonlinear dynamical system of the form
\begin{equation}\label{flow}
	\frac{d {\bf y}}{dt} = f({\bf y}, t)\,, \quad {\bf y}(0) = {\bf x} \in \mathbb{R}^{N_{s}}\,.
\end{equation}
We denote the associated flow of \eqref{flow} by ${\bf y}(t) = \varphi(t;{\bf x})$. The associated Hilbert space of observables is $L_{2}\left(\mathbb{R}^{N_{s}},\mathbb{R},\mu\right)$, for which we use the shorthand notation $L_{2}\left(\mathcal{O}\right)$, where a function $g: \mathbb{R}^{N_{s}} \longrightarrow \mathbb{R}$ satisfies $g \in L_{2}\left(\mathcal{O}\right)$ if
\[
	\int_{\mathbb{R}^{N_{s}}} \left|g({\bf x})\right|^{2} d\mu\left({\bf x}\right) < \infty
\]
for some appropriate measure $\mu$.  

One can gain considerable insight by examining the associated linear representation of the problem. This representation is given by the infinite-dimensional Koopman operator 
\[
	\mathcal{K}^{t}:L_{2}\left(\mathcal{O}\right)\rightarrow L_{2}\left(\mathcal{O}\right)\,. 
\]
For $g \in L_{2}\left(\mathcal{O}\right)$, we have that
\[
	\mathcal{K}^{t}g({\bf x}) =  g\left(\varphi(t;{\bf x})\right)\,.
\]
The power of moving to a linear-operator framework is that we capture the dynamics of the nonlinear system \eqref{flow}, as measured via observables, using the eigenvalues of $\mathcal{K}^{t}$.  We assume that $\mathcal{K}^{t}$ has a discrete spectrum.  If we can find a basis of $g \in L_{2}\left(\mathcal{O}\right)$ via the Koopman eigenfunctions $h_{j}$, which satisfy
\[
	\mathcal{K}^{t}h_{j} = e^{t\lambda_{j}}h_{j}\,,
\]
then it follows for any other observable $g$ that 
\begin{equation}
	g = \sum_{j=1}^{\infty}c_{j}h_{j}\,, \quad \mathcal{K}^{t}g = \sum_{j=1}^{\infty}e^{t\lambda_{j}}c_{j}h_{j}\,.
	\label{koopmanansatz}
\end{equation}

It is typically impossible to determine the modes of the Koopman operator $\mathcal{K}^{t}$ in closed form. Therefore, scholars have developed the dynamic-mode decomposition (DMD) \cite{schmid,mezic1, kutz} and extended dynamic-mode decomposition (EDMD)\cite{williams, williams2} to yield methods for the practical numerical computation of a finite number of the Koopman modes.  

To do DMD, we start by sampling the flow $\varphi(t;{\bf x})$ at a discrete times $t_{j}=t_{i} + (j-1)\delta t$ (with $j=1,\ldots, N_{T}+1$) to generate a data set ${\bf y}_{j} = \varphi(t_{j},{\bf x})$. If we select the set ${\bf g} = \{g_{l}\}_{l=1}^{M}$ of observables such that $g_{l}({\bf x}_{l}) = x_{l}$, then the DMD approximates $\mathcal{K}^{\delta t}$ by computing the spectra of the finite-dimensional operator $\tilde{K}_{a}$ that we obtain by solving the optimization problem 
\[
	{\bf K}_{a} = \mbox{arg min}_{{\bf K}}\|{\bf Y}_{+} - {\bf K}{\bf Y}_{-}\|_{F}^{2}\,,
\]
where ${\bf K}$ is an $N_{s}\times N_{s}$ matrix and 
\[
	{\bf Y}_{-} = \left[{\bf y}_{1} ~{\bf y}_{2} \cdots {\bf y}_{N_{T}} \right]\,, \quad {\bf Y}_{+} = \left[{\bf y}_{2} ~{\bf y}_{3} \cdots {\bf y}_{N_{T}+1} \right]\,.
\]
After obtaining the $N_{s}\times N_{s}$ matrix ${\bf K}_{a}$, we compute its singular-value decomposition (SVD) ${\bf K}_{a} = {\bf U}{\bf \Sigma} {\bf V}^{\dagger}$ \cite{carla2012}, where each factor is an $N_{s}\times N_{s}$ matrix, ${\bf U}$ and ${\bf V}$ are unitary, ${\bf \Sigma}$ is diagonal, and $^\dagger$ denotes the Hermitian conjugate.  Following the standard convention, we order the diagonal entries $\sigma_1, \ldots, \sigma_M$ (i.e., the singular values) of ${\bf \Sigma}$ so that 
\[
	\sigma_{1}\geq \sigma_{2}\geq \cdots \geq \sigma_{M}\geq 0 \,.  
\]
We then define the diagonal matrix $\tilde{{\bf \Sigma}}$ with diagonal entries
\begin{equation}	\label{dmdthresh}
	\tilde{\sigma}_{j} = \left\{
\begin{array}{rl}
	\sigma_{j}\,, & \log_{10}\left(\frac{\sigma_{j}}{\sigma_{1}} \right) \geq t_{\text{DMD}}\\
0\,, & \log_{10}\left(\frac{\sigma_{j}}{\sigma_{1}} \right) < t_{\text{DMD}}\,,
	\end{array}
\right.
\end{equation}
where we set the DMD threshold $t_{\text{DMD}}$ to remove singular values that are more reflective of ill-conditioning in ${\bf K}_{a}$ than of meaningful information. To be useful in practice, we take $t_{\text{DMD}} < 0$.  We then work with the matrix $\tilde{{\bf K}}_{a} = {\bf U}\tilde{{\bf \Sigma}}{\bf V}^{\dagger}$.  See \cite{rowley2} for a discussion of this DMD method and related approaches.  

We let $\tilde{{\bf K}}_{a} = {\bf \Xi} e^{\delta t {\bf \Lambda}} {\bf \Xi}^{-1}$ and write 
\[
	{\bf y}_{j} = \sum_{n=1}^{M}\xi_{n}e^{j\delta t \lambda_{n}} h_{n}({\bf x})\,, \quad h_{n}({\bf x}) = \left({\bf \Xi}^{-1}{\bf x}\right)_{n}\,.
\]
The real part of the eigenvalue $\lambda_{n}$ gives the amplitude of the $n$th mode, and the imaginary part of $\lambda_{n}$ gives its oscillation frequency. We determine the evolution of the associated eigenfunctions of the Koopman operator using the formula 
\[
	h_{n}({\bf y}_{j}) = \left({\bf \Xi}^{-1}{\bf y}_{j}\right)_{n}\,.
\]

A measure of error is the extent to which the computed modes $h_{n}({\bf x})$ behave as Koopman eigenfunctions \cite{kutz7}. To quantify this, we follow \cite{rowley} and calculate
\[
	\mathcal{E}_{j} = \frac{\sum_{n=1}^{N_{T}}\left|h_{j}\left({\bf y}_{n+1}\right)-e^{\delta t \lambda_{j}}h_{j}\left({\bf y}_{n}\right) \right|}{\sum_{n=1}^{N_{T}}\left|h_{j}\left({\bf y}_{n}\right)  \right|}
\]
for each $j \in \{ 1, \ldots, N_T+1\}$. The quantity $\mathcal{E}_{j}$ gives a normalized measure of how well the computed approximations to the Koopman eigenfunctions and eigenvalues are able to linearize the dynamics as the exact ones in \eqref{koopmanansatz}.  We choose a tolerance $\epsilon_{m}$ and keep only the modes that satisfy $\mathcal{E}_{j} < \epsilon_{m}$. We denote the number of the modes that we keep by $N_{r}$. We enforce how well these $N_{r}$ modes reconstruct the time series by choosing a reconstruction-error parameter $\epsilon_{rc}$ so that 
\[
	\frac{\|{\bf Y}_{+} - {\bf H}\|_{F}}{\|{\bf Y}_{+}\|_{F}} < \epsilon_{rc}\,,
\]
where $H$ is the matrix with columns
\[
	{\bf h}_{j} =  \sum_{l=1}^{N_{r}}\xi_{n_{l}}e^{j\delta t \lambda_{n_l}} h_{n_l}({\bf x})
\]
and $n_{l}$ denotes the subset of modes that satisfy the criterion $\mathcal{E}_{n_{l}}<\epsilon_{m}$.  

There is an interplay between the choices of $t_{\text{DMD}}$, $\epsilon_{m}$, and $\epsilon_{rc}$. It takes effort (e.g., through trial and error) to balance these parameters to produce meaningful results. For example, if $t_{\text{DMD}}$ is too small (e.g., $t_{\text{DMD}}=-16$, which corresponds to machine precision on most desktop computers), then one typically corrupts a DMD computation to the point that there is no practical way to find reasonable choices of $\epsilon_{m}$ or $\epsilon_{rc}$.  However, if $t_{\text{DMD}}$ is too large, only a very small number of modes meet the $\epsilon_{m}$ criterion, rendering it difficult to make reasonable choices of $\epsilon_{rc}$.  Likewise, setting $\epsilon_{m}$ to a value that is too small can produce excellent approximations to Koopman modes, but it is hard for the modal reconstruction to allow reasonable choices of $\epsilon_{rc}$. In Section \ref{sec4}, we describe parameter choices that reflect the necessary balancing.


\subsection{Coupled Oscillators on Random Graphs}

We describe a graph using an associated adjacency matrix ${\bf A}$ with elements $A_{jk}$. We assume that each of our graphs, which we assemble from random-graph models, are undirected, so their associated adjacency matrices are symmetric. We also assume that each graph has no self-edges (so $A_{jj}= 0$ for all $j$) and no multi-edges. Finally, we assume that our graphs are unweighted, so all entries of each matrix ${\bf A}$ are either $1$ or $0$. 

To study dynamics on a graph, suppose that each node $j \in \{1, \ldots, N_s\}$ is associated with a Kuramoto oscillator \cite{rodrigues}. This yields the following dynamical system: 
\begin{equation}
	\dot{\theta}_{j} = \omega_{j} + \frac{K}{N_{s}}\sum_{k=1}^{N_{s}}A_{jk}\sin(\theta_{k}-\theta_{j})\,, \quad \omega_{j} \sim \frac{1}{\gamma} g\left(\frac{x}{\gamma}\right)\,,
\label{kuramoto}
\end{equation}
where $\theta_j \in [0, 2\pi)$ is the phase of the $j$th oscillator, $\omega_j$ is the natural frequency of the $j$th oscillator, $K \geq 0$ controls the coupling strength between oscillators, and $g(y)$ is a mean probability distribution with mean $0$ and width $1$ (which constitutes a variance of $1$, provided the variance of the distribution $g$ is well-defined). Because of our rescaling, the parameter $\gamma$ is the variance of the distribution; we obtain identical oscillators in the limit $\gamma \rightarrow 0^{+}$.  

We are interested in the extent to which the oscillators {\it lock}. In the strongest sense, locking means {\it phase locking}, which is defined as 
\begin{equation}
	\lim_{t\rightarrow \infty}\theta_{j}(t) = \theta_{p}\,, \quad  j \in \{1, \ldots, N_s \}\,.
\label{phaselock}
\end{equation}
We also consider {\it frequency locking}, which is defined as
\[
	\lim_{t\rightarrow \infty}\dot{\theta}_{j}(t) = \omega_{f}\,, \quad j \in \{1, \ldots, N_s \}\,.
\]
In typical scenarios, phase locking implies frequency locking (although this need not be true in the presence of noise \cite{pimenova2016}), but the converse is not true in general.

To measure the extent that the oscillators lock, we calculate the order parameter  
\[
	r_{p}(t)e^{i\psi_{p}(t)} = \frac{1}{N_{s}}\sum_{j=1}^{N_{s}}e^{i\theta_{j}(t)}\,,
\]
and we note that $0\leq r_{p}(t) \leq 1$. If the oscillators are equally spaced at time $t$, such that $\theta_{j} = \frac{2\pi (j-1)}{N_{s}}$, then 
\[
	r_{p}(t) e^{i\psi_{p}(t)} = \frac{1}{N_{s}}\sum_{j=1}^{N_{s}}\left(e^{2\pi i/N_{s}}\right)^{(j-1)} = 0\,,
\]
so a value of $r_{p}(t)$ that is sufficiently close to $0$ for finite $N_s$ indicates an absence of phase locking between the oscillators. 

If we satisfy the phase-locking criterion \eqref{phaselock}, then $r_{p}(t)\rightarrow 1$ and 
\[
	r_{p}(t)e^{i\psi_{p}(t)} \rightarrow e^{i\theta_{p}} \,\, \text{as} \,\, t\rightarrow \infty\,.
\]
{For frequency locking}, it can be true that
\[
	\theta_{j}(t) \rightarrow \omega_{f}t + \theta_{j,s} \,\,\, \text{as} \,\,\,  t\rightarrow \infty\,,
\]
where $\theta_{j,s}$ is a phase shift.  In this case, 
\begin{equation}\label{eqn:freq_limit}
	r_{p}(t)e^{i\psi_{p}(t)} \rightarrow e^{i\omega_{f}t}\frac{1}{N_{s}}\sum_{j=1}^{N_{s}}e^{i\theta_{j,s}} \quad \text{as}\,\, t \rightarrow \infty\,.
\end{equation}
Depending on the particular locations of the phase shifts $\theta_{j,s}$, it is possible that $r_{p}(t)$ is small in magnitude. However, as one can see from Equation \eqref{eqn:freq_limit}, we also expect $r_{p}(t)$ to be almost constant in time.  We use practical criteria to determine when a finite set of oscillators are phase locked and/or frequency locked. We understand a collection of oscillators as being close to a phase-locked state when $r_{p}(t)$ is close to $1$ in magnitude and does not vary significantly in time. We understand a collection of oscillators as being close to a frequency-locked state if $r_{p}(t)$ is not too close to $0$ and does not vary significantly in time. 

Because we are characterizing {the closeness to} locking in an imprecise way, we do perturbation theory to facilitate the interpretation of our later numerical results. We perturb with respect to the mean $\left\langle r_{p}\right\rangle$ of $r_{p}(t)$ over the time interval $[t_{i},t_{f}]$. To compute this mean, we calculate
\begin{equation*}
	\left\langle r_{p} \right\rangle = \frac{1}{t_{f}-t_{i}}\int_{t_{i}}^{t_{f}}r_{p}(t)\, dt\,.  
\end{equation*}
For frequency locking, a natural condition for our perturbations is to suppose as $t$ becomes large that we can write each phase $\theta_{j}(t)$ in the form
\begin{equation}\label{eqn:slow_freq}
	\theta_{j}(t) = \omega_{f}t + \tilde{\theta}_{j}(\epsilon t)\,, \quad ~ 0< \epsilon \ll 1 \,.
\end{equation}
The ansatz in Equation \eqref{eqn:slow_freq} allows slow modulations around the locking frequency, where we control the extent of the slowness using the small positive parameter $\epsilon$.  We see that
\begin{align*}
	r_{p}(t) =  \left|\frac{1}{N_{s}}\sum_{j=1}^{N_{s}}e^{i\theta_{j}(t)} \right| 
		= \left|\frac{1}{N_{s}}\sum_{j=1}^{N_{s}}e^{i\tilde{\theta}_{j}(\epsilon t)} \right|\,,
\end{align*}
which demonstrates that $r_{p}(t)$ varies slowly with respect to time. We then write the order parameter as $r_{p}(t) = \tilde{r}_{p}(\epsilon t)$, where 
\[
	\tilde{r}_{p}(\epsilon t) = \left|\frac{1}{N_{s}}\sum_{j=1}^{N_{s}}e^{i\tilde{\theta}_{j}(\epsilon t)} \right|\,.
\]
Making the reasonable assumption that $\tilde{r}_{p}(\epsilon t)$ has a well-defined derivative $\dot{\tilde{r}}_{p}(\epsilon t)$, we then compare $r_{p}(t)$ to the mean for $t\in[t_{i},t_{f}]$ and write
\begin{align}
	r_{p}(t) - \left<r_{p} \right> &= \frac{1}{t_{f}-t_{i}}\int_{t_{i}}^{t_{f}}\left(\tilde{r}_{p}(\epsilon t)- \tilde{r}_{p}(\epsilon s)\right) ds \nonumber \\
&= \frac{1}{t_{f}-t_{i}}\int_{t_{i}}^{t_{f}}\left(\tilde{r}_{p}(\epsilon s + \epsilon (t-s))- \tilde{r}_{p}(\epsilon s)\right) ds \nonumber \\
&= \frac{\epsilon \dot{\tilde{r}}_{p}(\epsilon \xi)}{t_{f}-t_{i}}\int_{t_{i}}^{t_{f}}  (t-s) \, ds \quad (\text{with}\,\,\xi \in (t_{i},t_{f})\,) \nonumber \\
&= \dot{\tilde{r}}_{p}(\epsilon \xi)\left(\epsilon t - \epsilon \frac{\left(t_{i}+t_{f}\right)}{2} \right), \label{slowvar}
\end{align}
where we have used the mean-value theorem for the penultimate equality. This calculation shows that if oscillators vary slowly around a frequency-locked state, then the variation of the magnitude of the order parameter around the mean should (1) also be slow and (2) be bounded by the magnitude of the derivative multiplied by the difference between the time and the midpoint of the interval over which we average. As our numerical results demonstrate (see Section \ref{sec4}), frequency locking is more typical than phase locking in our networks of heterogeneous oscillators. Therefore, throughout the remainder of our paper, we make frequent recourse to Eq.~\eqref{slowvar}.  

Similarly, we claim that a reasonable perturbation around a phase-locked state is 
\begin{equation*}
	\theta_{j}(t) = \theta_{p} + \epsilon\tilde{\theta}_{j}(t)\,, \quad 0 < \epsilon \ll 1\,.
\end{equation*}
Using the Taylor-series expansion
\[
	e^{i\epsilon\tilde{\theta}_{j}} \approx 1 + i\epsilon \tilde{\theta}_{j} - \frac{\epsilon^{2}}{2}\tilde{\theta}^{2}_{j}
\]
yields
\begin{align}\label{smallvar}
	r_{p}(t) &\approx \left(1 + \epsilon^{2}\left( \tilde{m}^{2}(t)-\frac{1}{N_{s}}\sum_{j=1}^{N_{s}}\tilde{\theta}^{2}_{j}(t)\right)\right)^{1/2} \nonumber \\
&\approx 1 + \frac{\epsilon^{2}}{2}\left( \tilde{m}^{2}(t)-\frac{1}{N_{s}}\sum_{j=1}^{N_{s}}\tilde{\theta}^{2}_{j}(t)\right) \nonumber \\
&\approx 1 - \frac{\epsilon^{2}}{2N_{s}}\sum_{j=1}^{N_{s}}\left(\tilde{\theta}_{j}(t) - \tilde{m}(t) \right)^{2}\,, 
\end{align}
where 
\[
	\tilde{m}(t) = \frac{1}{N_{s}}\sum_{j=1}^{N_{s}}\tilde{\theta}_{j}(t)\,.
\]
In the second line of \eqref{smallvar}, we used the Taylor-series approximation $\sqrt{1+x^{2}}\approx 1 + x^{2}/2$. From Eq.~\eqref{smallvar}, we can now see clearly how variances around the mean of the perturbations of the locked phase $\theta_{p}$ reduce the magnitude of the order parameter $r_{p}(t)$ below $1$ and introduce temporal variations. We also anticipate that if the perturbations $\tilde{\theta}_{j}$ vary slowly, then $r_{p}(t)$ should also vary slowly.


The issues that arise from the above choice of order parameter have inspired efforts, such as the use of topological data analysis \cite{topaz2015}, to apply data-analysis techniques to study order parameters in complex systems.  Moreover, in addition to the global order parameter that we employ, one can calculate other order parameters to examine localized locking in subsets of oscillators~\cite{arenas2007sync,buzna2009sync,sedina}. Applying a DMD approach with these alternative order parameters is worthwhile to explore in future research efforts.

Additionally, as was shown in \cite{restrepo}, one can derive analytical criteria that guarantee phase locking of some subset of Kuramoto oscillators on a network. Specifically, in the asymptotic limit of infinitely many oscillators oscillating and arbitrarily long times, there is necessarily some locking of some subset of the oscillators if 
\[
	\gamma < \gamma_{c}\,, \quad \gamma_{c} = \frac{\pi N_{s} K g(0) \sigma^{{\bf \text{A}}}_{\text{max}}}{2}\,,
\]
where $\sigma^{{\bf \text{A}}}_{\text{max}}$ is the largest eigenvalue of the adjacency matrix ${\bf A}$.  However, the number of oscillators that lock (or the number of communities of oscillators that lock) is not determined by this criterion.  Moreover, in the present paper, we consider systems with a relatively small number of oscillators, so we are far away from the asymptotic regime that was studied in \cite{restrepo}. 


\section{Estimation of Locking and Functional-Community Detection Using the Dynamic-Mode Decomposition} \label{sec3}

To measure the amount of locking among the oscillators in the time interval $[t_{i},t_{f}]$, we examine the collection $\left\{\xi_{n}\right\}_{n=1}^{N_{r}}$ of Koopman modes. We define the {\it overlap matrix} ${\bf C}^{\text{o}}$ with elements 
\[
	C^{\text{o}}_{jl} = \left|\sum_{n=1}^{N_{r}}\hat{\xi}_{n,j}\hat{\xi}^{\ast}_{n,l}\right|\,, \quad \hat{\xi}_{n} = \frac{\xi_{n}}{\|\xi_{n}\|_{2}}\,.
\]
Using the Cauchy--Schwarz inequality, we prove that 
\[
	0\leq C^{\text{o}}_{jl} \leq 1 \,.
\]
We also see that ${\bf C}^{\text{o}}$ is symmetric. Using a threshold of $C_{\text{cr}}\in (0,1)$, we construct graphs of the strongest interactions by generating an associated adjacency matrix ${\bf A}^{(\text{md})}(t_{i},t_{f})$ with elements
\[
	A^{(\text{md})}_{jl}(t_{i},t_{f}) = \left\{
\ba{rl}
1\,, &  C^{\text{o}}_{jl} \geq C_{\text{cr}} \\ 
\\
0\,, & C^{\text{o}}_{jl} < C_{\text{cr}}\,.
\ea
	\right.  
\] 
Because ${\bf C}^{\text{o}}$ is symmetric, it is also true that ${\bf A}^{(\text{md})}(t_{i},t_{f})$ is symmetric and thus that its associated graph is undirected. In a given time interval, we define subsets of the oscillators to belong to a ``community'' if they are a part of the same connected component of this graph.
 
We expand on this notion of community to track the merging, separation, formation, and dissolution of communities over time. This adds then to a growing body of literature on detecting communities in time-varying networks; see, for example, \cite{palla2007,mucha, bassett2013, holme}. We separate a fixed time interval $[t_{i},t_{f}]$ into $n_{w}$ subintervals $\left\{I_{j}\right\}_{j=1}^{n_{w}}$ of equal length. We also generate $n_{w}-1$ equally spaced intervals $\left\{\tilde{I}_{j}\right\}_{j=1}^{n_{w}-1}$ that connect the midpoints of the $I_{j}$ intervals. We order these intervals to give a total of $2n_{w}-1$ intervals $\bar{I}_{j}$ such that 
\[
	\bar{I}_{j} = \left\{\ba{rl} 
I_{(j+1)/2}\,, & 1 \equiv j ~\mbox{mod} ~2 \\ 
\tilde{I}_{j/2}\,, & 0 \equiv j ~\mbox{mod} ~2 \,.\\ 
\ea
	\right.  
\]
With this construction, our collection of intervals overlaps in time. We generate a sequence of communities using the sequence
\[
	\left\{{\bf A}^{(\text{md})}(\bar{I}_{j})\right\}_{j=1}^{2n_{w}-1}  
\]
of adjacency matrices. We use overlapping time intervals to help ensure some continuity of the communities from one interval to the next. See \cite{porter4} for further exploration of this issue.  

To study how communities evolve over time, we define a time-dependent graph $\mathcal{G}_{t_{i},t_{f}}$ that tracks community relationships across time. To generate this graph, for the first time interval $\bar{I}_{1}$, we use ${\bf A}^{(\text{md})}(\bar{I}_{1})$ to obtain communities $\left\{C_{k}(\bar{I}_{1})\right\}_{k=1}^{N_{c,1}}$ of oscillators in which each community corresponds to one connected component of the graph that is associated with ${\bf A}^{(\text{md})}(\bar{I}_{1})$.  Each of these communities corresponds to an individual node in $\mathcal{G}_{t_{i},t_{f}}$. After initializing $\mathcal{G}_{t_{i},t_{f}}$, we generate successive graphs in a sequence using the following recursive process. Assuming that we have constructed $\mathcal{G}_{t_{i},t_{f}}$ up to the $j^{\text{th}}$ time step, we place an edge between the $k^{\text{th}}$ community $C_{k}(\bar{I}_{j})$ from ${\bf A}^{(\text{md})}(\bar{I}_{j})$ and the $l^{\text{th}}$ community $C_{l}(\bar{I}_{j+1})$ from ${\bf A}^{(\text{md})}(\bar{I}_{j+1})$ if the nodes in $C_{k}(\bar{I}_{j})$ are identical to or are a subset of those in $C_{l}(\bar{I}_{j+1})$.  Otherwise, if the nodes in $C_{k}(\bar{I}_{j})$ are not contained within a community or have fractured into smaller communities at the $(j+1)^{\text{th}}$ time step, then $C_{k}(\bar{I}_{j})$ terminates at the $j^{\text{th}}$ time. This process ensures that $\mathcal{G}_{t_{i},t_{f}}$ is a forest that adaptively connects communities to each other across time based on the relative overlaps of Koopman modes. In our numerical computations (see Section \ref{sec4}), we observe coalescence of communities when there is phase locking or frequency locking.

In summary, our community-detection algorithm to generate $\mathcal{G}_{t_{i},t_{f}}$ proceeds as follows:
\begin{algorithm}[H] \label{algo}
Given a time series $\left\{{\bf y}_{n}\right\}_{n=1}^{N_{s}}$ in the time interval $[t_{i},t_{f}]$ 
\begin{algorithmic}
\Require Choose $C_{\text{cr}}$ and $n_{w}$. Separate the interval $[t_{i},t_{f}]$ into overlapping subintervals $\left\{\bar{I}_{j}\right\}_{j=1}^{2n_{w}-1}$.
\For{$j \in \{1, \ldots, 2n_{w} - 1\}$}
\begin{enumerate} 
\item Compute the DMD of the portion of the time series that is in the interval $\bar{I}_{j}$.  
\item Compute the corresponding overlap matrix ${\bf C}^{\text{o}}$ and its associated adjacency matrix $A^{(\text{md})}(\bar{I}_{j})$ using the threshold $C_{\text{cr}}$.
\item Compute the associated communities $\left\{C_{k}(\bar{I}_{j})\right\}_{k=1}^{N_{c,j}}$. For each of the $N_{c,j}$ communities, add a node to $\mathcal{G}_{t_{i},t_{f}}$.
\end{enumerate}
\If{$j > 1$}
\For{$k \in \{1, \ldots, N_{c,j}\}$}
\For{$l \in \{1, \ldots, N_{c,j-1}\}$}
\If {$C_{k}(\bar{I}_{j})\subseteq C_{l}(\bar{I}_{j-1})$} 
\State Form an edge between the corresponding nodes in $\mathcal{G}_{t_{i},t_{f}}$.
\State Break
\EndIf
\EndFor
\EndFor
\EndIf
\EndFor
\end{algorithmic}
\caption{Threshholding algorithm for functional community detection.}
\end{algorithm}

\section{Numerical Experiments} \label{sec4}

We now conduct a variety of numerical experiments to examine how successfully DMD can detect functional communities in networks of Kuramoto oscillators. Our networks have $N_{s}=800$ oscillators, and the coupling parameter is $K=10$. We examined networks with as many as $N_{s}=3200$ oscillators, and we obtain similar results to the ones that we will discuss below.  We draw the natural frequencies of the oscillators from a Gaussian distribution, so $\gamma_{c}=4000\sqrt{\pi}\sigma^{{\bf \text{A}}}_{\text{max}}$. Given this choice of distribution, we expect that at least subset of oscillators will always become at least nearly locked because $\sigma^{{\bf \text{A}}}_{\text{max}} > 1$ in all of our experiments.  We use a second-order Runge--Kutta solver with a step size of $\delta t = 0.05$ to integrate the Kuramoto model \eqref{kuramoto}. We run all simulations up to $t_{f}=800$. From $t = 760$ to $t_{f}=800$, we perform DMD on the circle around which the oscillators move, so we use the coordinates $\left(\cos(\theta_{j}(t)),\sin(\theta_{j}(t))\right)$.  

We set the DMD threshold in Eq.~\eqref{dmdthresh} to $t_{\text{DMD}}=-4$, the error threshold of the DMD modes to $\epsilon_{m}=10^{-1}$, and the reconstruction error to $\epsilon_{\text{rc}}=10^{-1}$. As we described at the end of Section \ref{dmd}, these parameter choices reflect a balance that we obtain through trial and error. For our numerical experiments, we report results for heterogeneity parameters $\gamma = 0.1$, $\gamma = 1$, and $\gamma = 10$. Although smaller values of $\gamma$ (indicating more homogeneous oscillators) allow greater viable ranges for the three DMD parameters, our choices represent values that work well across the two-decade range of the values of $\gamma$.

In our simulations, we use two classical random-graph models (RGMs): the preferential-attachment (PA) model of Barab\'{a}si--Albert (BA) \cite{barabasi} and the Watts--Strogatz (WS) small-world model \cite{watts}. After specifying the parameter values of an RGM, we generate a single graph from the model. When we simulate the Kuramoto oscillators on either network, we initialize the dynamical system with $N$ oscillators that we space uniformly around the circle. In the following paragraphs, we indicate the specific variants and parameter values that we present for these models. We also examined other parameter choices for these RGMs, and we obtained similar results.

For our BA network, we start with a seed network of $n = 10$ nodes with edges that we choose according to a $G(n,p)$ Erd\H{o}s--Renyi (ER) model with an independent, uniform probability of $p = 0.75$ for placing an edge between nodes. At each time step, we add a new node to the network; this node has $4$ edges that we connect uniformly at random (without replacement) to existing nodes. We grow our BA network until it consists of $N = 800$ nodes. 

Our WS network also has $N = 800$ nodes. We start with a ring lattice in which each node is adjacent to $K_{\text{nbh}} = 20$ nearest neighbors.  We then rewire the network as follows. Starting at the $j^{\text{th}}$ node $n_{j}$, we consider half of its nearest-neighbor nodes, where the specific neighbors $n_l$ are those with the indices
\[
	l=\tilde{l} ~ \text{mod} ~ N\,, \quad j < \tilde{l} \leq j+\frac{K_{\text{nbh}}}{2}\,.
\]
For each $n_{l}$, with probability $\beta$, we do the following.
\begin{enumerate}
\item We choose a node $n_{k}$ uniformly at random from nodes that are not adjacent to $n_{j}$.
\item We remove the edge $e_{jl}$ that connects $n_{j}$ to $n_{l}$.
\item We add an undirected edge $e_{jk}$ to connect $n_{j}$ to $n_{k}$.
\end{enumerate}
As we iterate through the nearest neighbors of $n_j$, we do not allow rewiring to any nodes that currently or formerly are nearest neighbors. However, such a nearest-neighbor connect can arise again when we rewire connections from other nodes after we have moved on from $n_j$. Therefore, it is possible for nearest-neighbor edges to be removed and added back again, although this is not common. In our simulations, we set the rewiring probability to $\beta = 0.6$. As discussed in \cite{newman-book2018}, the rewiring process drastically reduces the mean geodesic distance between nodes from that of the initial ring graph. 

In our computations, we study the structure of the largest-connected component (LCC) in the adjacency matrix ${\bf A}^{(\text{md})}([760, 800])$. Our method of detecting functional communities always generates some output (regardless of whether it is interpretable), and it is instructive to examine the details of the predominant community that we obtain in the final examined time interval. We thereby improve our understanding of the meaning of the communities that we obtain using our approach. To help illustrate the structure of the LCC, we compare the local clustering coefficients of the LCC to those in the original RGM. {Note that all of our subsequent calculations with the oscillators in the LCC use the time-series output of our original simulation.  This yields our order parameters\footnote{Note that we do not compute the order parameters of the LCC by simulating Kuramoto dynamics using only oscillators that are in the LCC.}.

For the $j^{\text{th}}$ node of a network, we calculate the local clustering coefficient \cite{newman-book2018}
\[
	C_{t}(j) = \frac{\mbox{number of triangles through node} ~j}{\bp d_{j} \\ 2\ep}\,,
\]
where $d_{j}$ denotes the degree of node $j$. In Figure \ref{fig:fullgraphcluster}, we plot the local clustering coefficients of each node in networks that we construct using the BA and WS models. For most nodes, $C_{t}(j)$ is not particularly large for either RGM. Therefore, we do not expect to observe particularly modular structural communities in either RGM. In the WS model, the local clustering coefficients tend to be small because the rewiring probability is large.

\begin{figure}[!h]
\centering
\begin{tabular}{cc}
\includegraphics[width=.23\textwidth]{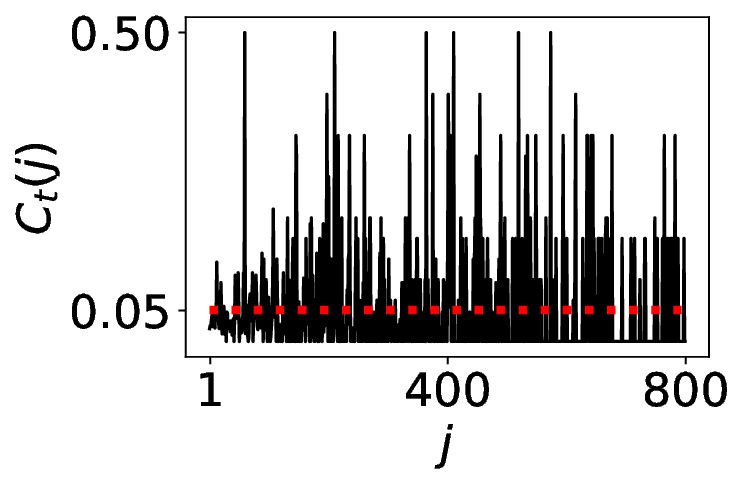} & \includegraphics[width=.23\textwidth]{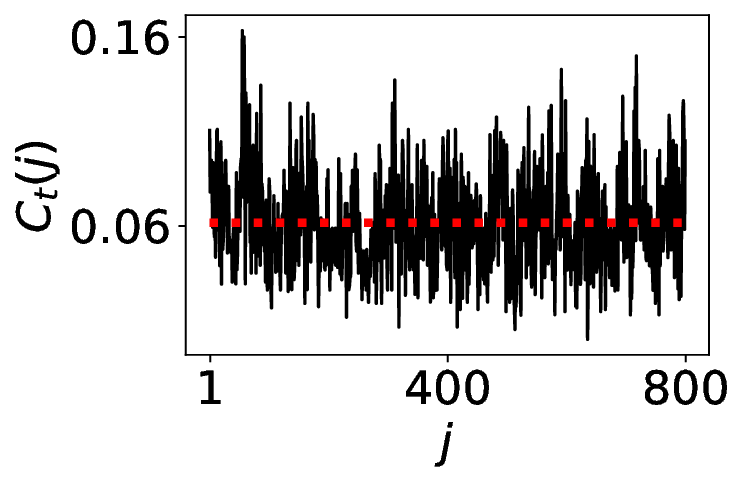}\\
\qquad(a) & \qquad(b) 
\end{tabular}
\caption{The local clustering coefficient $C_{t}(j)$ of each node $j$ for (a) a Barab\'{a}si--Albert (BA) network and (b) a Watts--Strogatz (WS) network.  The horizontal line is the mean value of $C_{t}(j)$. Neither of these networks tends to have particularly large clustering coefficients, although the mean local clustering coefficient is larger in the WS network than in the BA network because of the nature of the associated RGMs.}
\label{fig:fullgraphcluster}
\end{figure}

To generate communities such as the LCC, we set the threshold of the correlation graph to be $C_{\text{cr}} = 0.99$, except for the WS network with heterogeneity parameter $\gamma=10$, where we use $C_{\text{cr}}=0.98$. We do this because the threshold $C_{\text{cr}}=0.99$ only yields one or two oscillators in the LCC, which makes it difficult to measure locking. Using a slightly smaller value of $C_{\text{cr}}$ in this case ameliorates the issue and yields more intelligible results.  Among the values of $C_{\text{cr}}$ that we examine, these parameter choices typically produce the best community discrimination across the examined values of $\gamma$. Smaller values of $C_{\text{cr}}$ generate fewer (and larger) communities; at large values of $\gamma$, this can allow too much heterogeneity within a community, and we then observe little to no locking. For example, using $C_{\text{cr}}=0.95$ yields very similar results as $C_{\text{cr}} = 0.99$ when $\gamma=0.1$, but it yields communities with much less locking than for $C_{\text{cr}} = 0.99$ when $\gamma=1$ or $\gamma=10$ in the BA model. Our observations for the WS mode are similar. Arguably, it may be desirable to allow the value of $C_{\text{cr}}$ to change to adjust to the known heterogeneity (as quantified by $\gamma$) of the oscillators and produce communities with the most locking, but this increases the difficulty of comparing the functional communities that we obtain for different values of $\gamma$. Therefore, we use the value $C_{\text{cr}} = 0.99$ for most of our case studies to facilitate exposition and consistency in reporting results.  

Because we know about a key measurable aspect (which, in our case, is locking of the oscillators) of the underlying model, we are able to adjust our parameter choices to account for that phenomenon.  We expect our approach to be useful for problems in which the dynamics includes a similarly recognizable and quantifiable feature. There are a wide variety of ways in which one can define and identify such features in what one can term loosely as the task of ``coherent-structure identification''.  See \cite{haller2015,allshouse2015,holmes_book} for discussions of coherent structures in fluids and in other dynamical systems. Without such features, our DMD approach is likely to produce arbitrary results that are difficult to analyze and verify, and we would not suggest using our method in such situations.  

It is also worth considering other RGMs and multiple instances of an RGM for the same parameter values. We performed calculations using stochastic block models \cite{newman-book2018} and obtained results that are similar to those that we report in the present paper. If one considers multiple instances of an RGM, one can then average over the ${\bf C}^{\text{o}}$ matrices to obtain a consensus matrix and proceed to do community detection using the consensus matrix. This is an interesting question that is worth exploring in future work.


\subsection{Weakly Heterogeneous Oscillators: $\gamma = 0.1$ }

We first consider oscillators that are only weakly heterogeneous by examining the case $\gamma=0.1$. As one can see in Figures \ref{fig:omatrices_gam_ptone}(a,b), in which we compare the distribution of oscillator frequencies $\dot{\theta}_{j}$ at $t=0$ and $t=t_{f}=800$, the mostly homogeneous natural frequencies of the oscillators lends the system to frequency locking. In these figures, we also observe differences that arise from the different network topologies of WS and BA networks. There is only a small amount of locking in the BA network, whereas the WS network is in an almost fully locked state. The similarity in the oscillators and relatively long simulation times in both networks results in a relatively modest number of DMD modes [see Figure \ref{fig:omatrices_gam_ptone}(c,d)], with only about ten modes in each case.

Using ${\bf A}^{(\text{md})}$, in Figure \ref{fig:cluster_gam_ptone}(a,b), we show the sizes of the communities that we generate with a given correlation-graph threshold. In both the BA and WS networks, the LCC is the largest community by far and almost all other oscillators belong to their own single-oscillator communities. 
As expected from the locking dynamics, the LCC is much smaller in the BA network than in the WS network. In Figures \ref{fig:cluster_gam_ptone}(c,d), we see that the mean local clustering coefficient in the BA network is smaller than that in the WS network. However, in both networks, the values of the local clustering coefficient are markedly larger in the LCC of ${\bf A}^{(\text{md})}$ than in the original adjacency matrices of these networks. Compare Figures \ref{fig:cluster_gam_ptone}(c,d) to Figures \ref{fig:fullgraphcluster}(a,b).

\begin{figure}[!h]
\centering 
\begin{tabular}{cc}
\includegraphics[width=.23\textwidth]{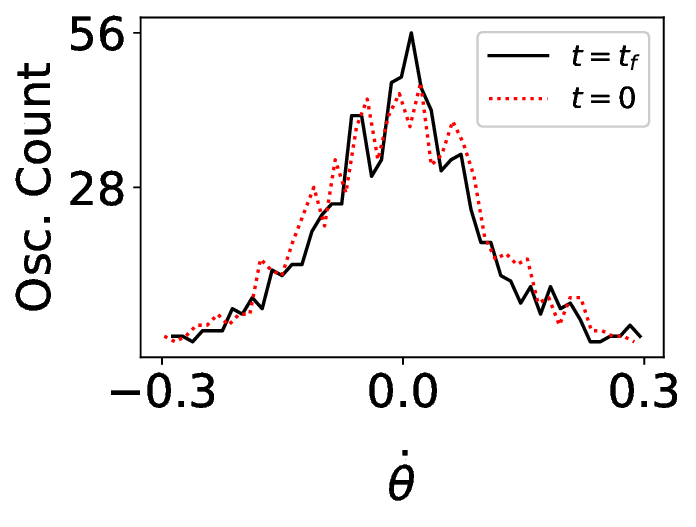} & \includegraphics[width=42mm, height=31mm]{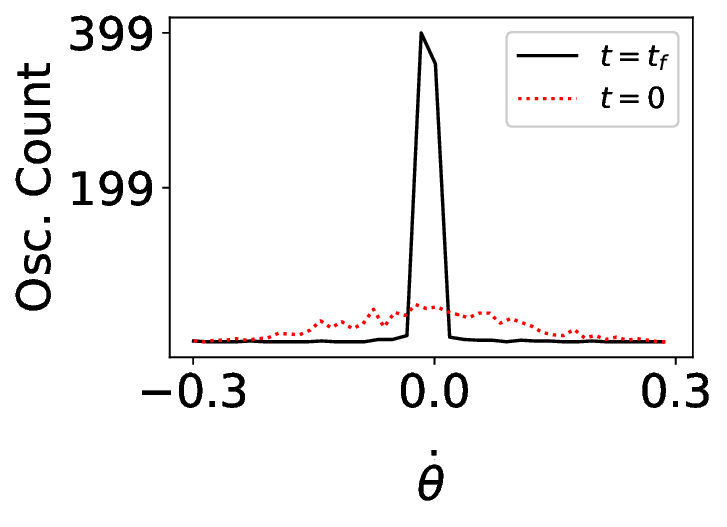}\\
\qquad(a) & \qquad(b) \\
\includegraphics[width=.23\textwidth]{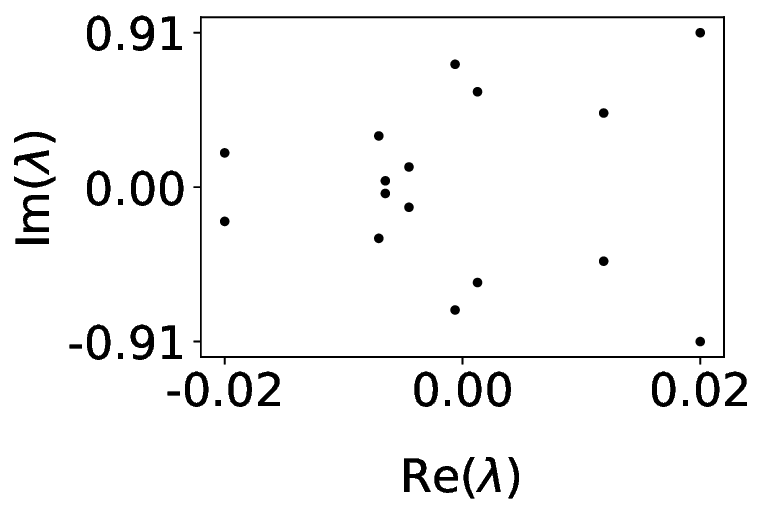} & 
\includegraphics[width=.23\textwidth]{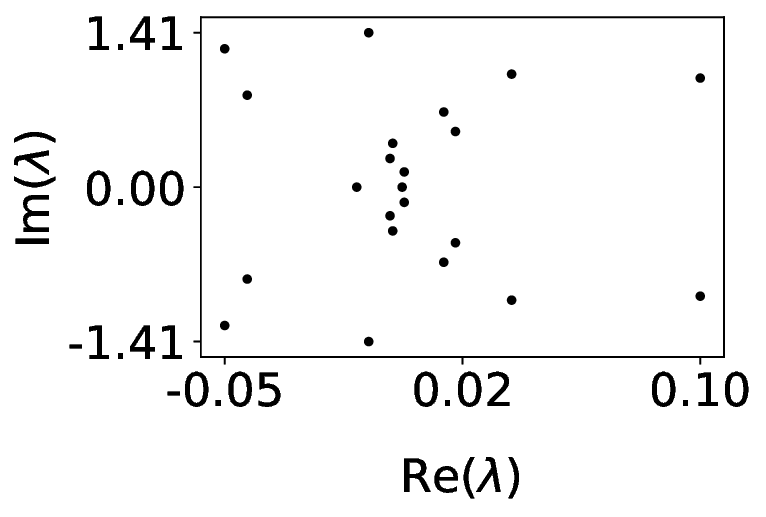} \\
\qquad(c) & \qquad(d)
\end{tabular}
\caption{(a,b) Frequency distributions and (c,d) DMD spectra for weakly heterogeneous Kuramoto oscillators (i.e., when $\gamma=0.1$). We show our results for a BA network in the left column and for a WS network in the right column. In (a,b), we plot the distributions of $\dot{\theta}_{j}$ at $t=0$ and $t=t_{f}$. Observe that the WS network clearly exhibits locking, whereas the BA network does not have a significant shift in the initial oscillator distribution. In (c,d), we show the associated DMDs of these networks. The BA network has a smaller frequency range than the WS network, as one can see in the values of $\text{Im}(\lambda)$. Additionally, the larger range of $\text{Re}(\lambda)$ values in the WS network than in the BA network indicate that the former experiences larger changes in amplitude.  In (a) and (b), ``Osc." stands for ``Oscillator".
}
\label{fig:omatrices_gam_ptone}
\end{figure}

\begin{figure}[!h]
\centering 
\begin{tabular}{cc}
\includegraphics[width=.23\textwidth]{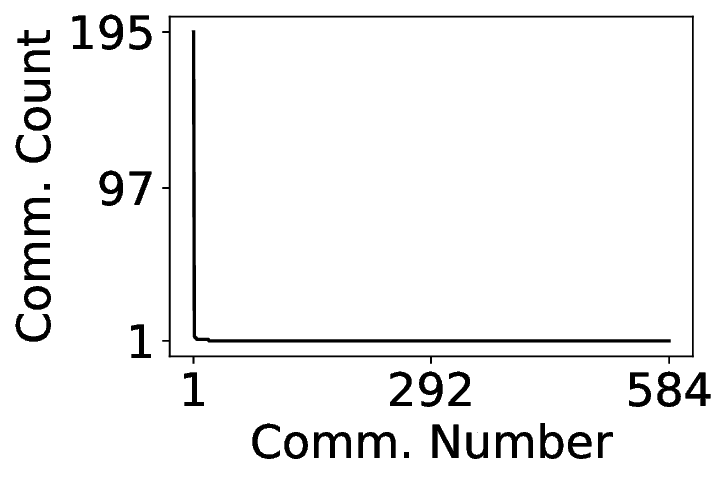} & 
\includegraphics[width=.23\textwidth]{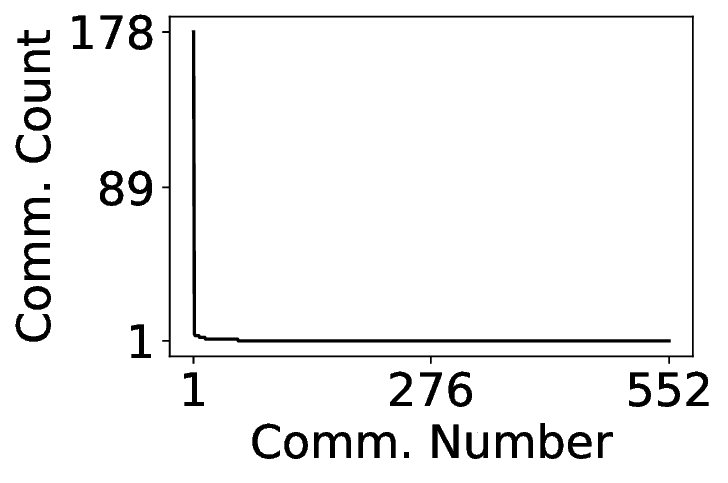} \\
\qquad(a) & \qquad(b) \\
\includegraphics[width=.23\textwidth]{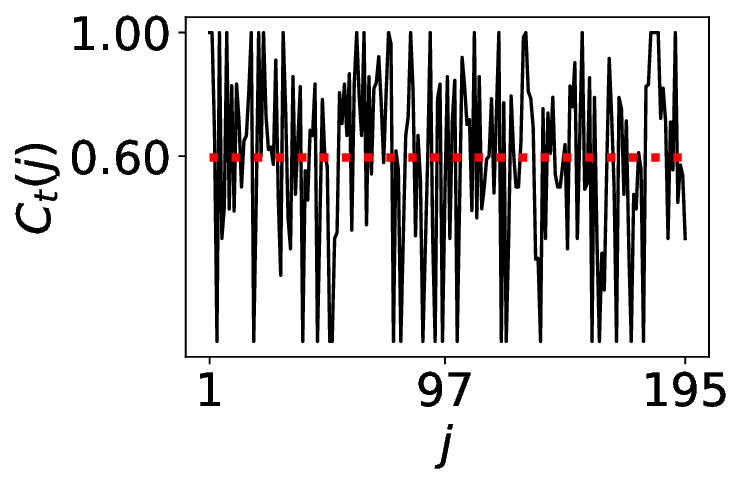} & 
\includegraphics[width=.23\textwidth]{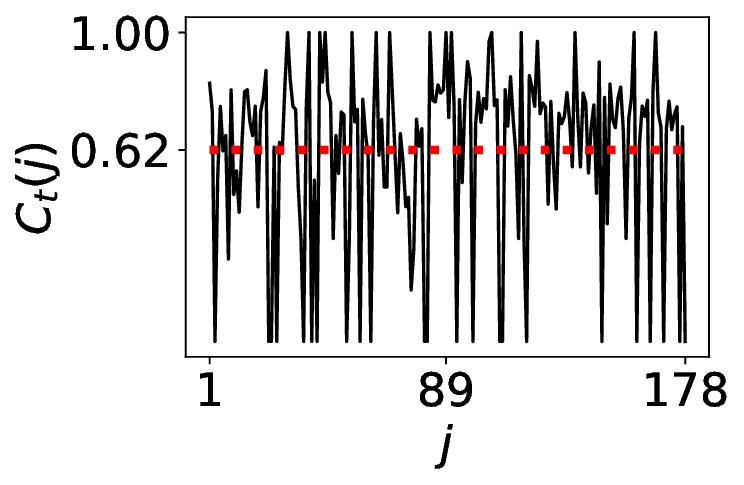} \\
\qquad(c) & \qquad(d)
\end{tabular}
\caption{(a,b) The numbers of nodes in each connected component of ${\bf A}^{(\text{md})}$ and (c,d) the local clustering coefficients of the nodes in the LCC of ${\bf A}^{(\text{md})}$ of the BA and WS networks for oscillators with $\gamma = 0.1$. We show our results for a BA network in the left column and for a WS network in the right column. In both the BA network and the WS network, the LCC is the largest community by far. Additionally, in both networks, the local clustering coefficients of the nodes in the LCC of ${\bf A}^{(\text{md})}$ are markedly larger than the baseline values in the original networks in Figure \ref{fig:fullgraphcluster}.  In (a) and (b), ``Comm." stands for ``Community".
}
\label{fig:cluster_gam_ptone}
\end{figure}

In Figure \ref{fig:order_param_gam_ptone}, we plot the order parameters for both the original networks and the associated LCCs (of ${\bf A}^{(\text{md})}$) of the BA and WS networks.  We see in Figures \ref{fig:order_param_gam_ptone}(e,f) (which we obtain after applying the thresholding from
Algorithm 1) that the order-parameter magnitude $r_{p}(t)$ oscillates less around the mean than in Figures \ref{fig:order_param_gam_ptone}(a,b).  By comparing the real parts of the phases of the order parameters in Figures \ref{fig:order_param_gam_ptone}(c,g), we see that the LCC of the BA network has stronger frequency locking than the original network. By contrast, given that the original WS network is almost fully phase locked, its LCC only exhibits a little bit more phase locking [see Figures \ref{fig:order_param_gam_ptone}(d,h)].

\begin{figure}[!h]
\centering 
\begin{tabular}{cc}
\includegraphics[width=.23\textwidth]{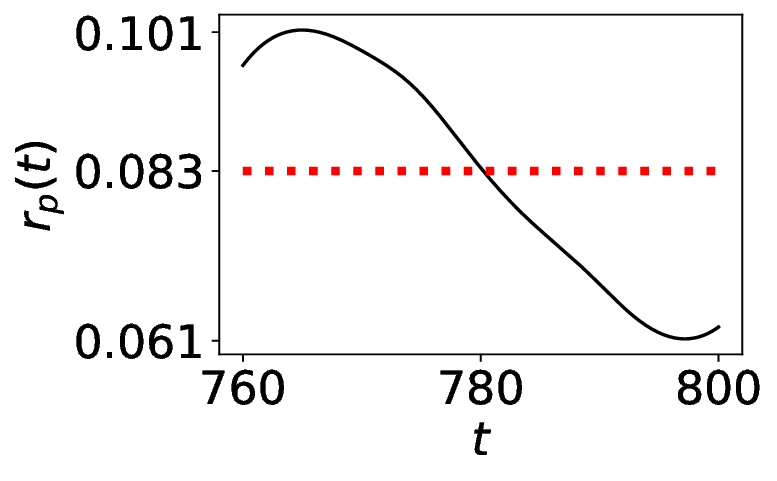} &  
\includegraphics[width=.23\textwidth]{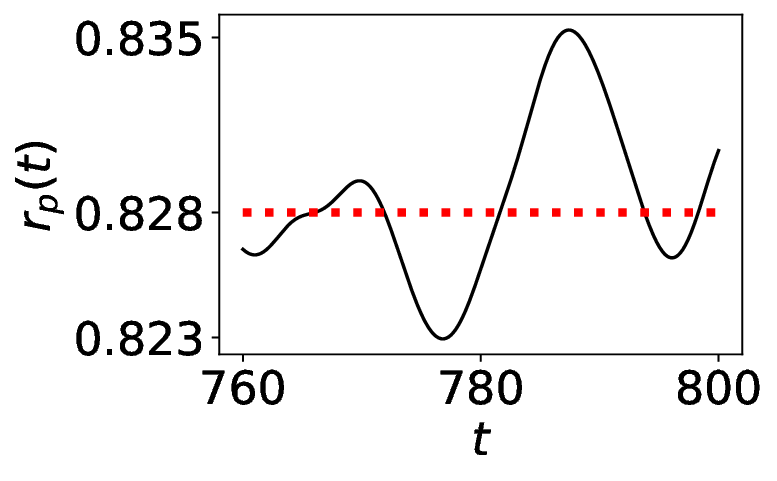}\\
\qquad(a) & \qquad(b) \\
\includegraphics[width=.23\textwidth]{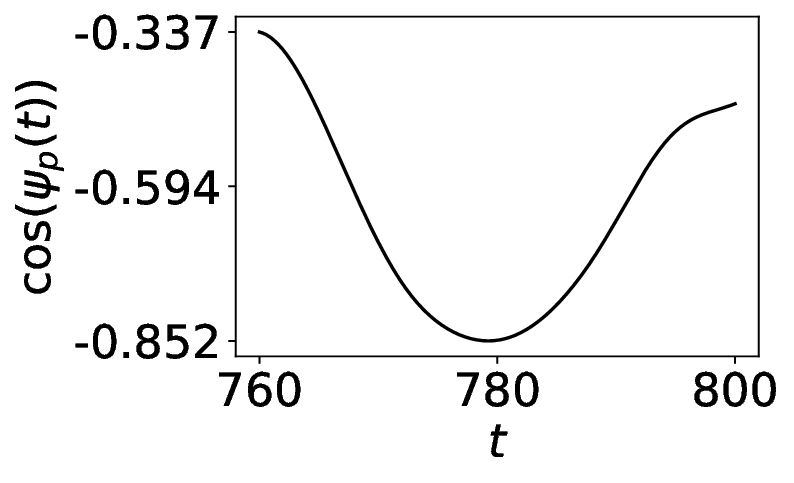} &  
\includegraphics[width=.23\textwidth]{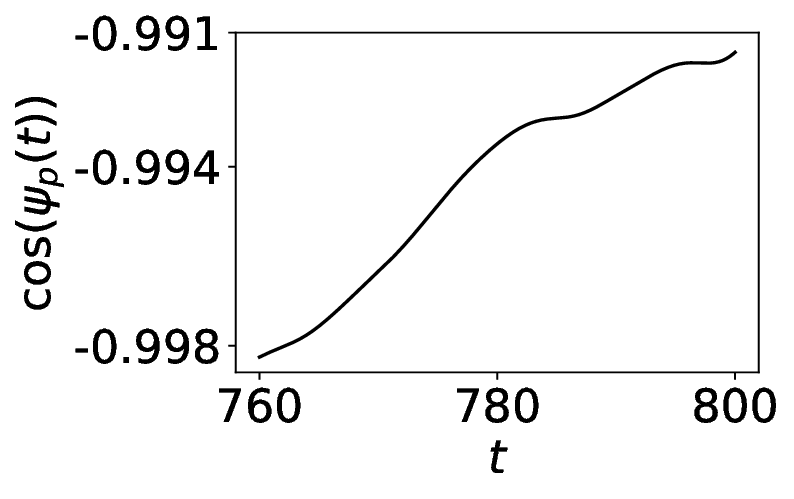}\\
\qquad(c) & \qquad(d) \\
\includegraphics[width=.23\textwidth]{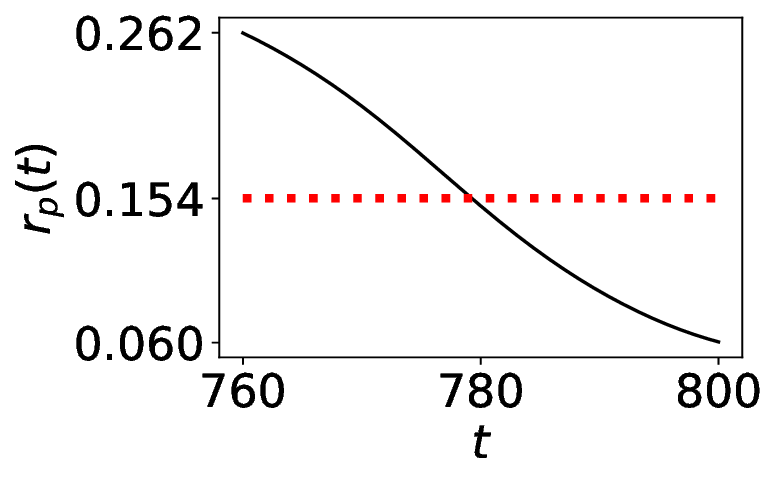}& 
\includegraphics[width=.23\textwidth]{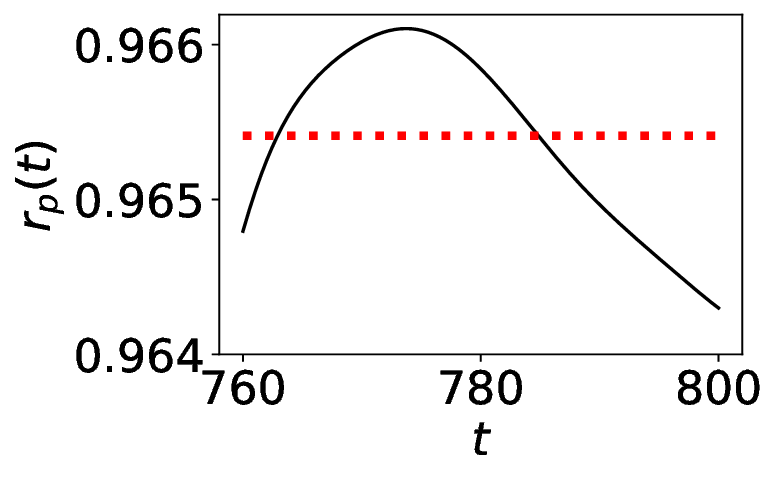}\\
\qquad(e) & \qquad(f) \\ 
\includegraphics[width=.23\textwidth]{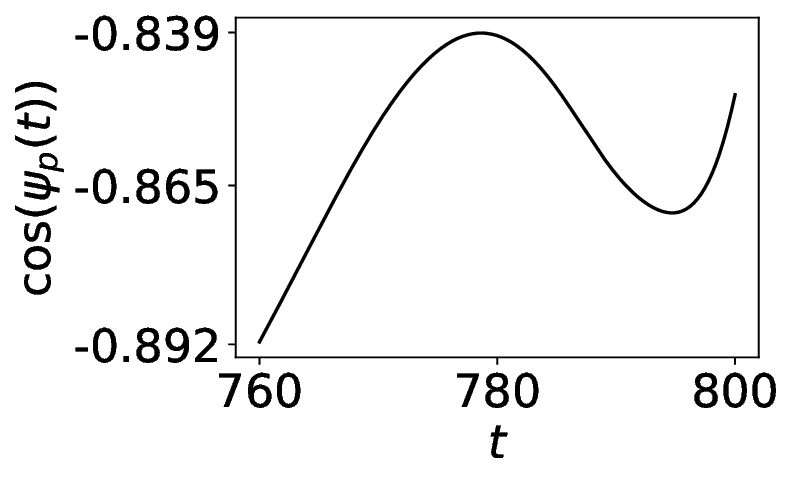}& 
\includegraphics[width=.23\textwidth]{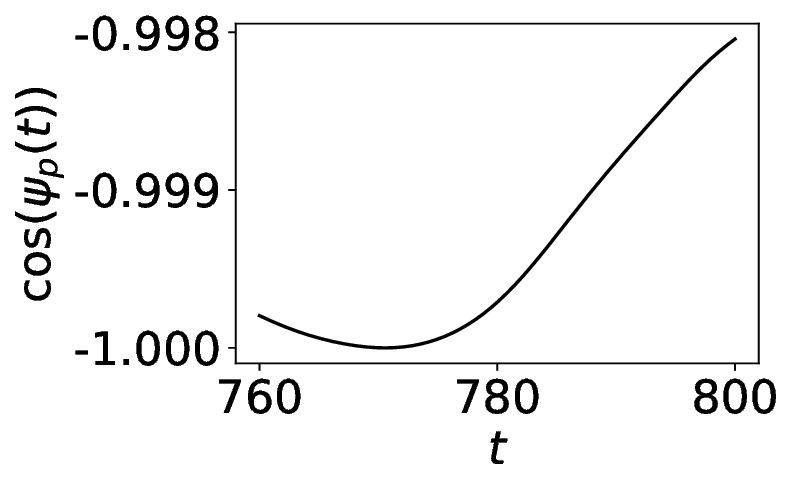}\\
\qquad(g) & \qquad(h) 
\end{tabular}
\caption{The (a,b,e,f) order-parameter magnitude $r_{p}(t)$ and (c,d,g,h) real part $\cos(\psi_{p}(t))$ of the order-parameter phase for the BA and WS networks with weakly heterogeneous oscillators (i.e., when $\gamma=0.1$). We show our results for a BA network in the left column and for a WS network in the right column. Our results for the original two networks are in panels (a)--(d), and our results for the LCCs of ${\bf A}^{(\text{md})}$ of these networks are in panels (e)--(h). The horizontal line in (a,b,e,f) indicates the mean $\left\langle r_{p}(\cdot)\right\rangle$ of the order-parameter magnitude. For the BA model, we observe stronger frequency locking in the LCC of ${\bf A}^{(\text{md})}$ than in the original network because the former has slower variations in $r_{p}(t)$ around the mean and more regular behavior of the phase. In the WS model, we see evidence that the LCC of ${\bf A}^{(\text{md})}$ has more phase locking than in the original network.}
\label{fig:order_param_gam_ptone}
\end{figure}


\subsection{Moderately Heterogeneous Oscillators: $\gamma = 1$ }

As we anticipate and confirm in our numerical computations, the small variance in the distribution of frequencies $\omega_{j}$ when $\gamma=0.1$ is amenable to the formation of locked states. The situation is starkly different when $\gamma=1$ and the natural frequencies of the oscillators are much more heterogeneous. As we can see in Figures \ref{fig:omatrices_gam_one}(a,b), there does not appear to a locked state on the same time scales that we observed when $\gamma=0.1$. Additionally, the DMD spectra in Figures \ref{fig:omatrices_gam_one}(c,d) now have about 50 modes, indicating a far greater complexity in the dynamics when $\gamma=1$ than when $\gamma=0.1$. This, in turn, results in much smaller functional communities [see Figures \ref{fig:cluster_gam_one}(a,b)]. Nevertheless, for both $\gamma=0.1$ and $\gamma=1$, the local clustering coefficients of the LCC of the ${\bf A}^{(\text{md})}$ networks are much larger than those in the original networks.

\begin{figure}[!h]
\centering 
\begin{tabular}{cc}
\includegraphics[width=.23\textwidth]{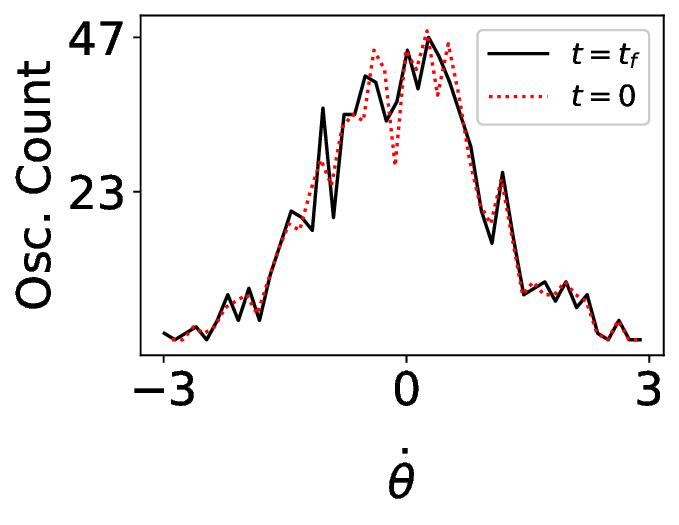} & 
\includegraphics[width=.23\textwidth]{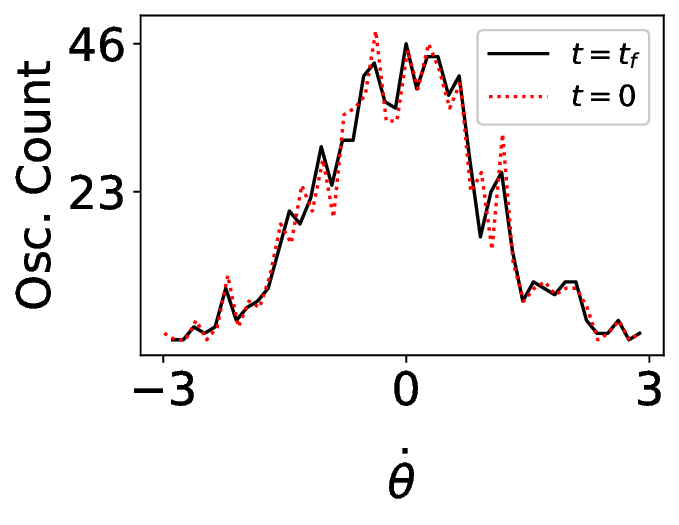} \\
\qquad(a) & \qquad(b) \\
\includegraphics[width=.23\textwidth]{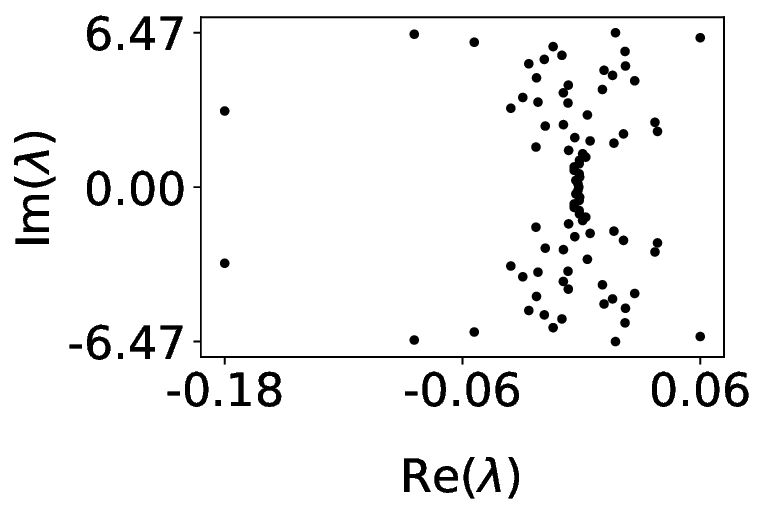} & 
\includegraphics[width=.23\textwidth]{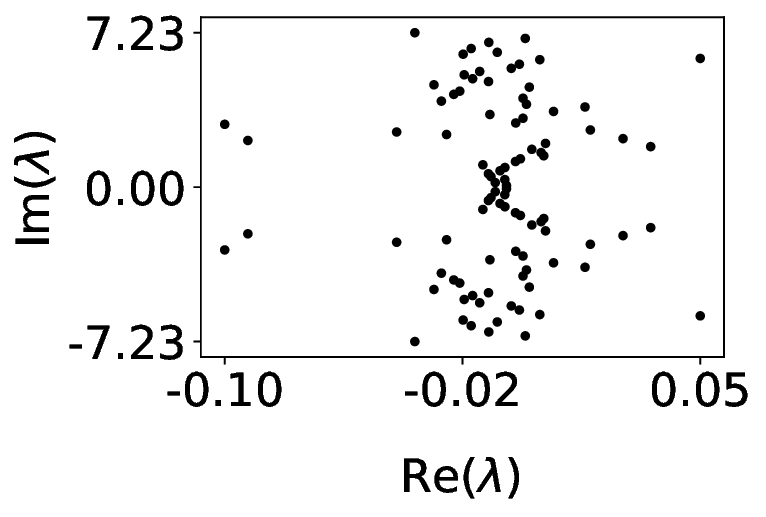} \\
\qquad(c) & \qquad(d) 
\end{tabular}
\caption{(a,b) Frequency distributions and (c,d) DMD spectra for moderately heterogeneous Kuramoto oscillators (i.e., when $\gamma=1$).  We show our results for a BA network in the left column and for a WS network in the right column. In (a,b), we plot the distributions of $\dot{\theta}_{j}$ at $t=0$ and $t=t_{f}$ for (a) a BA network and (b) a WS network. When $\gamma  = 1$, the characteristics of the networks from the two RGMs are much less distinguishable than when $\gamma=0.1$. However, the DMD spectra still have some differences, such as how eigenvalues cluster on the imaginary axis.  In (a) and (b), ``Osc." stands for ``Oscillator".}
\label{fig:omatrices_gam_one}
\end{figure}

\begin{figure}[!h]
\centering 
\begin{tabular}{cc}
\includegraphics[width=42mm, height=29mm]{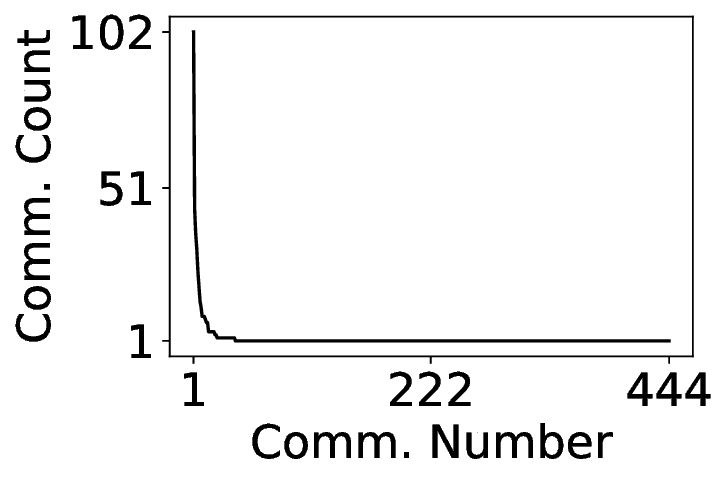} & 
\includegraphics[width=.23\textwidth]{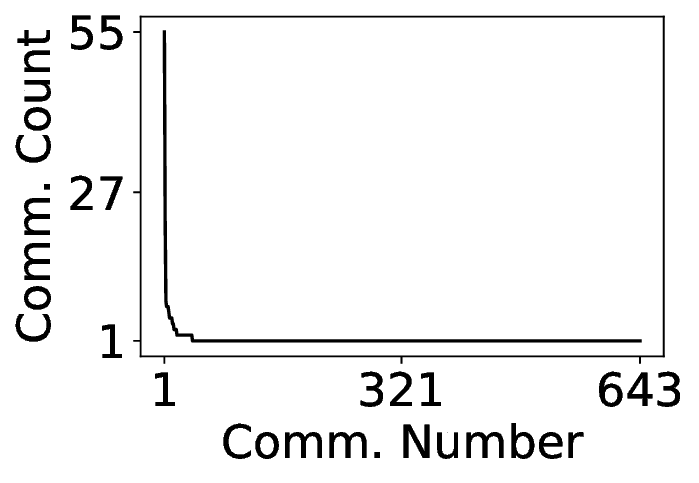} \\
\qquad(a) & \qquad(b) \\
\includegraphics[width=.23\textwidth]{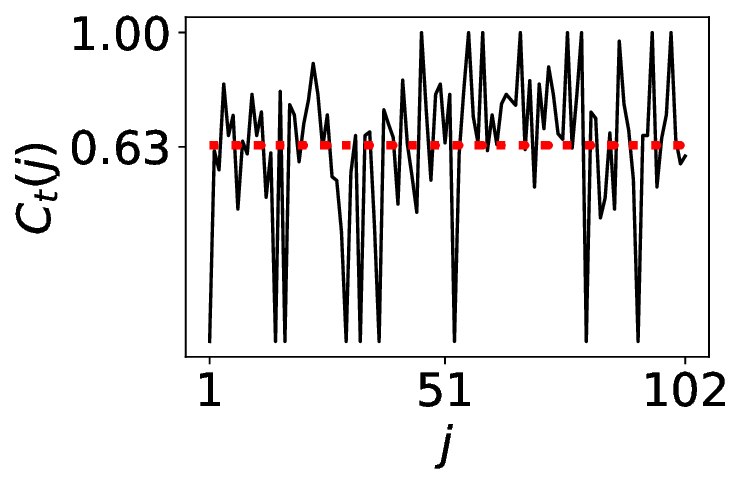} & 
\includegraphics[width=.23\textwidth]{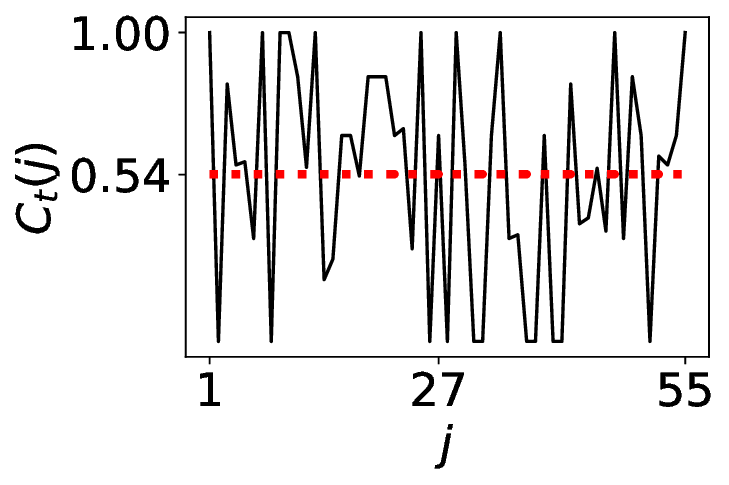} \\
\qquad(c) & \qquad(d) 
\end{tabular}
\caption{(a,b) The numbers of nodes in each connected component of ${\bf A}^{(\text{md})}$ and (c,d) the local clustering coefficients of the nodes in the LCC of ${\bf A}^{(\text{md})}$ of the BA and WS networks for oscillators with $\gamma = 1$. We show our results for a BA network in the left column and for a WS network in the right column. The LCCs are still markedly larger than any other community, but they are much smaller than they are when $\gamma = 0.1$. In both networks, the local clustering coefficients of the nodes in the LCC of ${\bf A}^{(\text{md})}$ are again markedly larger than the baseline values in the original networks in Figure \ref{fig:fullgraphcluster}, although this is now less extreme than when $\gamma = 0.1$.  In (a) and (b), ``Comm." stands for ``Community".}
\label{fig:cluster_gam_one}
\end{figure}

This clear absence of locking is echoed in Figures \ref{fig:order_param_gam_one}(a,b), which reveal that the order-parameter magnitudes $r_{p}(t)$ are markedly more oscillatory when $\gamma = 1$ than they are when $\gamma=0.1$. Additionally, we do not observe steady oscillations in $\cos(\psi_{p}(t))$ [see Figures \ref{fig:order_param_gam_one}(c,d)]. 
We see in Figures \ref{fig:order_param_gam_one}(e,f) that in the LCC of ${\bf A}^{(\text{md})}$, the oscillations of $r_{p}(t)$ around the mean are
smaller than those in the original networks.
Using Eq.~\eqref{slowvar}, we attribute this observation to the associated much stronger frequency locking of the oscillators in the LCC.
The far steadier and more uniformly oscillating phases in Figures \ref{fig:order_param_gam_one}(g,h) than in Figures \ref{fig:order_param_gam_one}(c,d)
corroborates this observation.


\begin{figure}[!h]
\centering 
\begin{tabular}{cc}
\includegraphics[width=.23\textwidth]{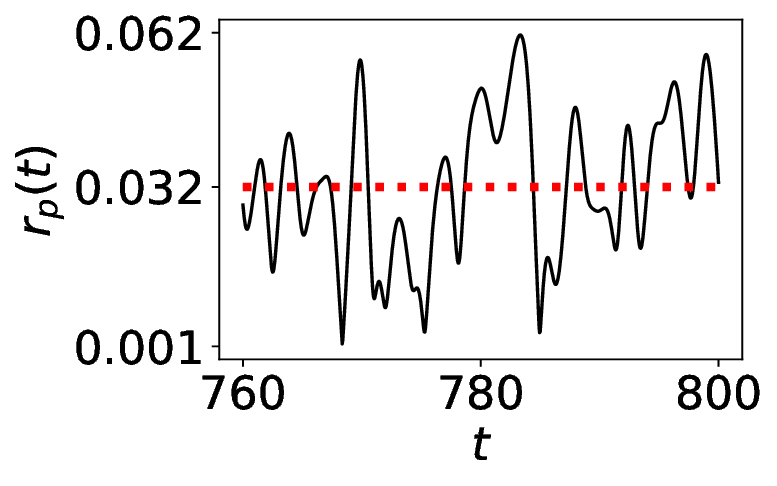} &  
\includegraphics[width=.23\textwidth]{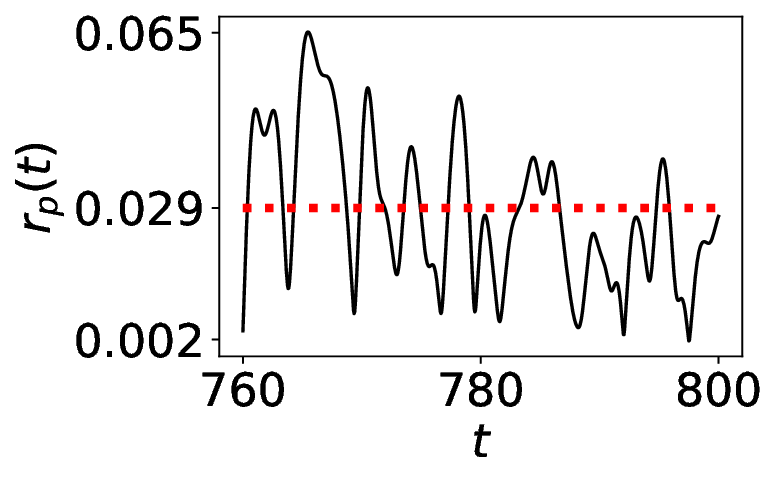}\\
\qquad(a) & \qquad(b) \\ 
\includegraphics[width=.23\textwidth]{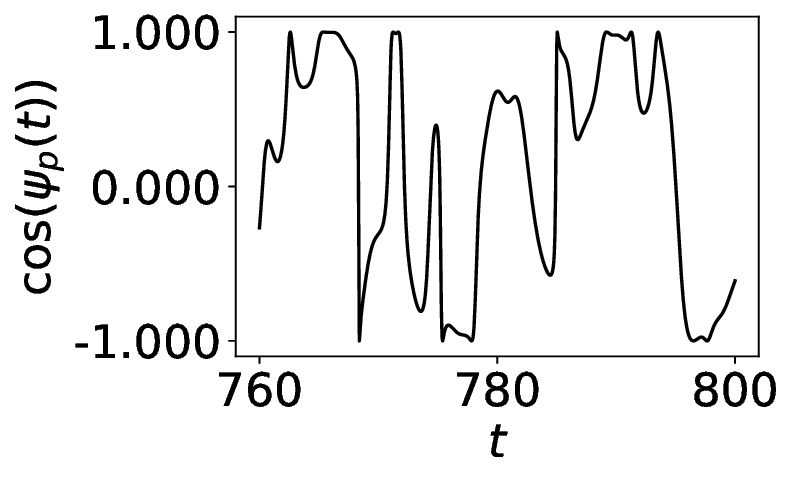} &  
\includegraphics[width=.23\textwidth]{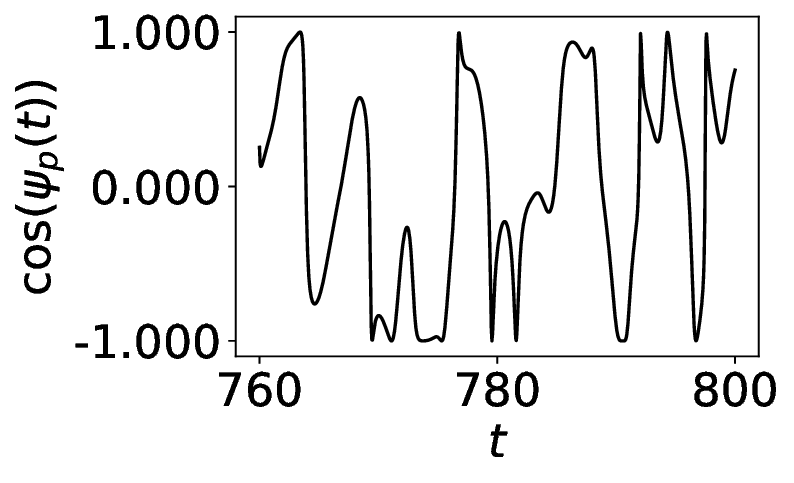}\\
\qquad(c) & \qquad(d) \\
\includegraphics[width=.23\textwidth]{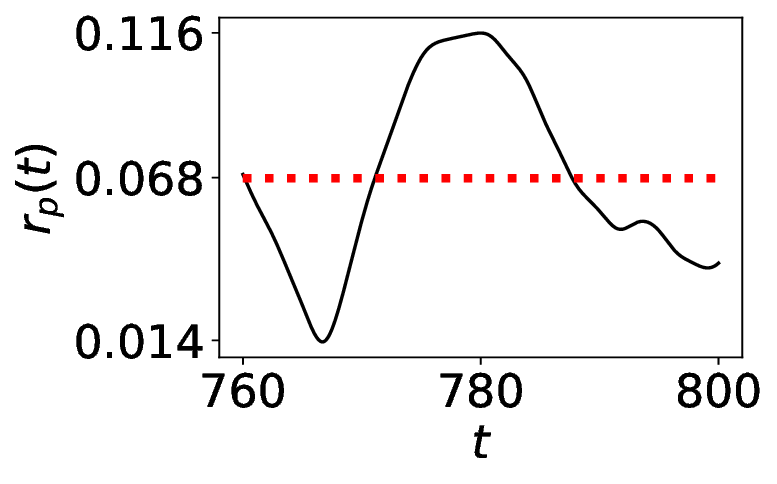}& 
\includegraphics[width=.23\textwidth]{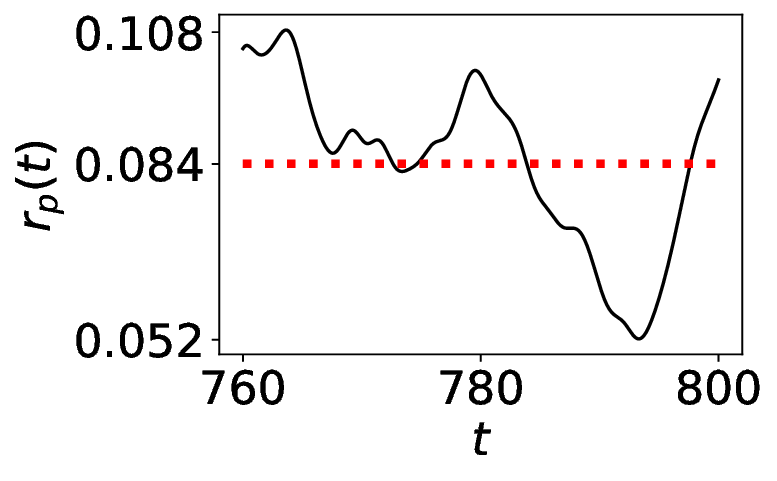}\\
\qquad(e) & \qquad(f) \\ 
\includegraphics[width=.23\textwidth]{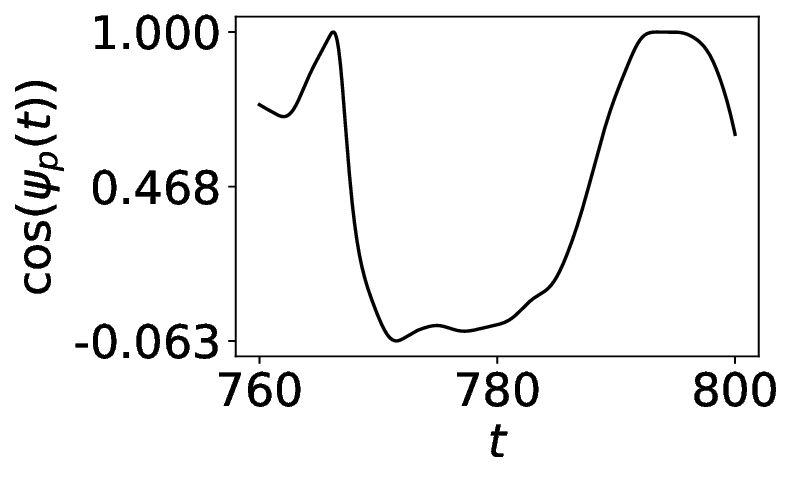}& 
\includegraphics[width=.23\textwidth]{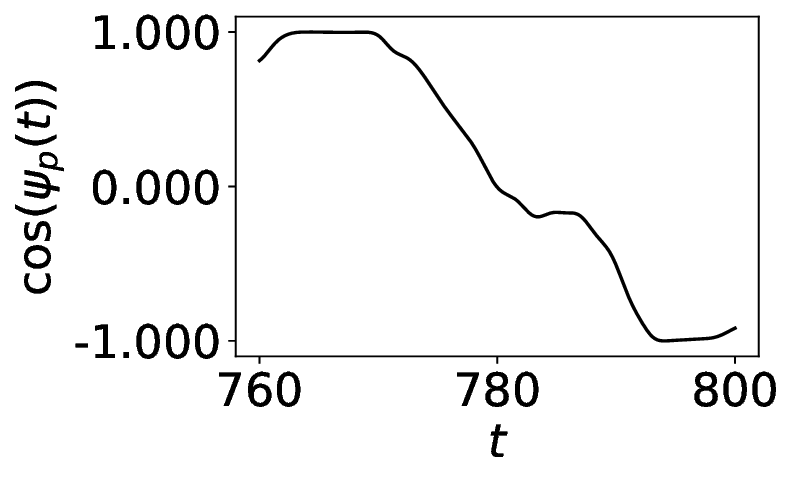}\\
\qquad(g) & \qquad(h) 
\end{tabular}
\caption{The (a,b,e,f) order-parameter magnitude $r_{p}(t)$ and (c,d,g,h) real part $\cos(\psi_{p}(t))$ of the order-parameter phase for the BA and WS networks when $\gamma=1$. Our results for the original two networks are in panels (a)--(d), and our results for the LCCs of ${\bf A}^{(\text{md})}$ of these networks are in panels (e)--(h). We show our results for a BA network in the left column and for a WS network in the right column. The horizontal line in (a,b,e,f) indicates the mean $\left\langle r_{p}(\cdot)\right\rangle$ of the order-parameter magnitude. For both the BA network and the WS network, the LCC of ${\bf A}^{(\text{md})}$ has much stronger frequency locking [as one can see by the slow variation in both $r_{p}(t)$ and $\cos(\psi_{p}(t))$] than in the original networks.
}
\label{fig:order_param_gam_one}
\end{figure}


\subsection{Strongly Heterogeneous Oscillators: $\gamma=10$}

Finally, we examine strongly heterogeneous oscillators by setting $\gamma=10$. Recall that we use a threshold of $C_{\text{cr}} = 0.99$ for the BA model and a threshold of $C_{\text{cr}} = 0.98$ for the WS model. We use a smaller threshold for the WS model to ensure that there are enough oscillators in the LCC to obtain meaningful measurements. Based on our previous results, we anticipate that we are unlikely to observe locking among any significant fraction of the oscillators on the time scales that we examine. Our computations in Figure \ref{fig:omatrices_gam_ten} confirm this expectation. 

In Figures \ref{fig:omatrices_gam_ten}(a,b), we see that there is almost no change in the frequency distributions during our simulations. This yields a complicated DMD spectra, with approximately 340 eigenvalues for each network, in Figures \ref{fig:omatrices_gam_ten}(c,d). The community counts in Figures \ref{fig:cluster_gam_ten}(a,b) reveal that even the LCCs of ${\bf A}^{(\text{md})}$ only have 7 oscillators for each network. However, as one can see in Figures \ref{fig:cluster_gam_ten}(c,d), these LCCs still have larger local clustering coefficients than the original networks.     

\begin{figure}[!h]
\centering 
\begin{tabular}{cc}
\includegraphics[width=.23\textwidth]{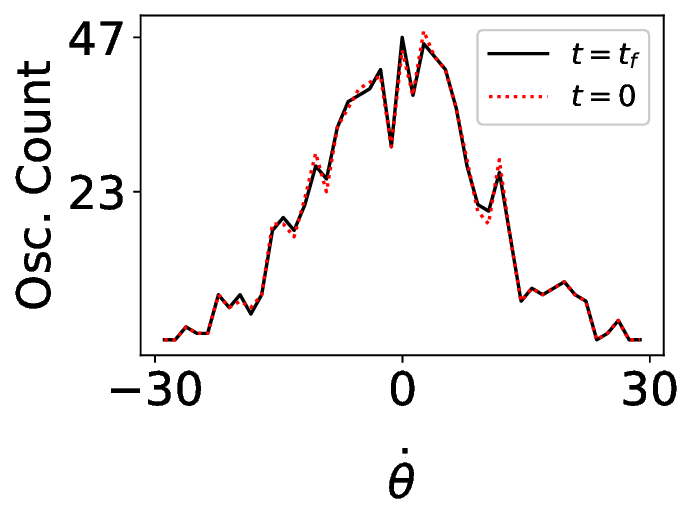} & 
\includegraphics[width=.23\textwidth]{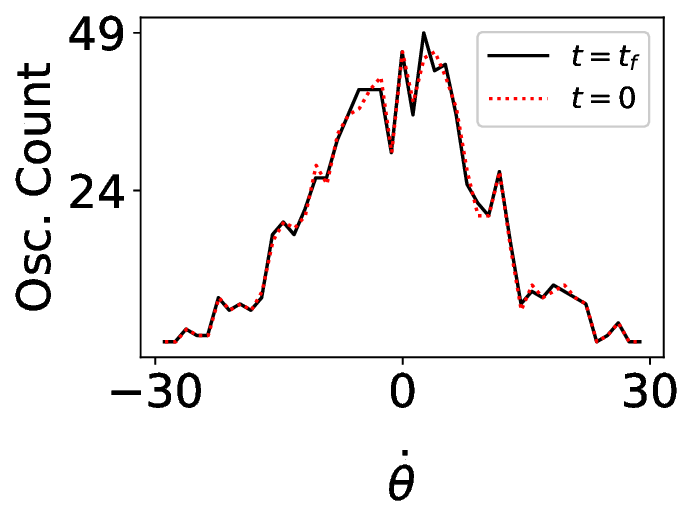} \\
\qquad(a) & \qquad(b) \\
\includegraphics[width=.23\textwidth]{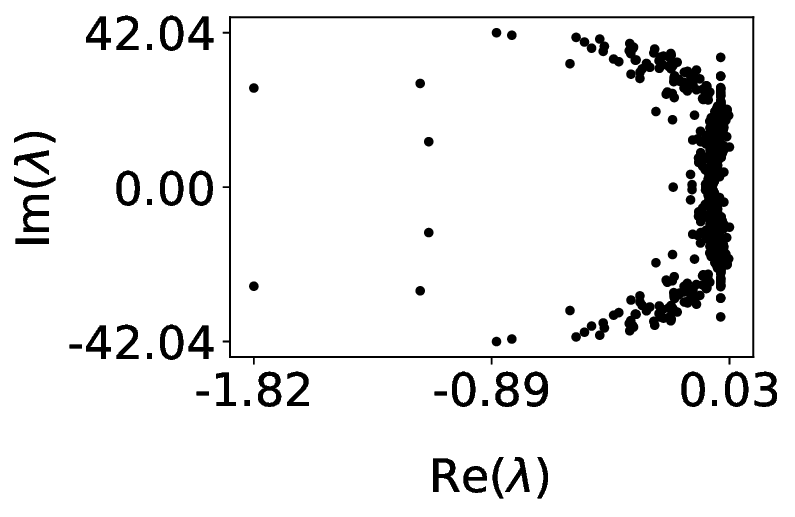} & 
\includegraphics[width=.23\textwidth]{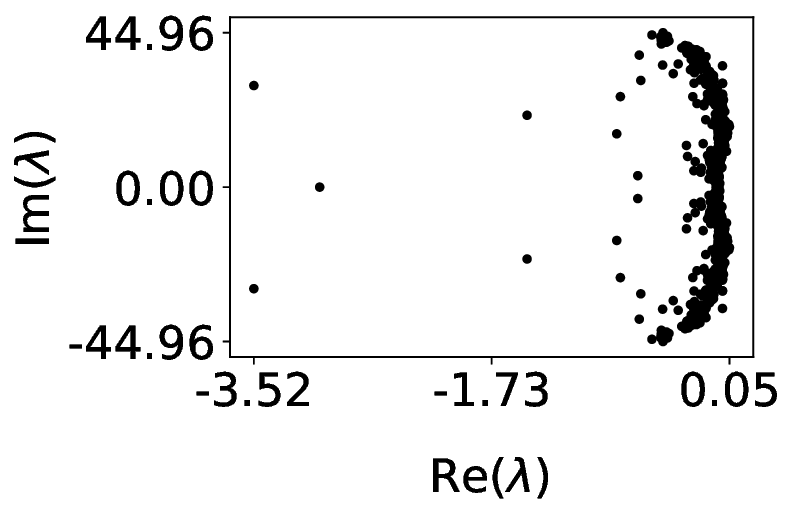} \\
\qquad(c) & \qquad(d)
\end{tabular}
\caption{(a,b) Frequency distributions and (c,d) DMD spectra for strongly heterogeneous Kuramoto oscillators (i.e., when $\gamma=10$). We show our results for a BA network in the left column and for a WS network in the right column. In (a,b), we plot the distributions of $\dot{\theta}_{j}$ at $t=0$ and $t=t_{f}$ for (a) a BA network and (b) a WS network. It is now difficult to distinguish between the networks from the two models, although we still observe slight differences in the DMD spectra.  In (a) and (b), ``Osc." stands for ``Oscillator".}
\label{fig:omatrices_gam_ten}
\end{figure}

\begin{figure}[!h]
\centering 
\begin{tabular}{cc}
\includegraphics[width=.23\textwidth]{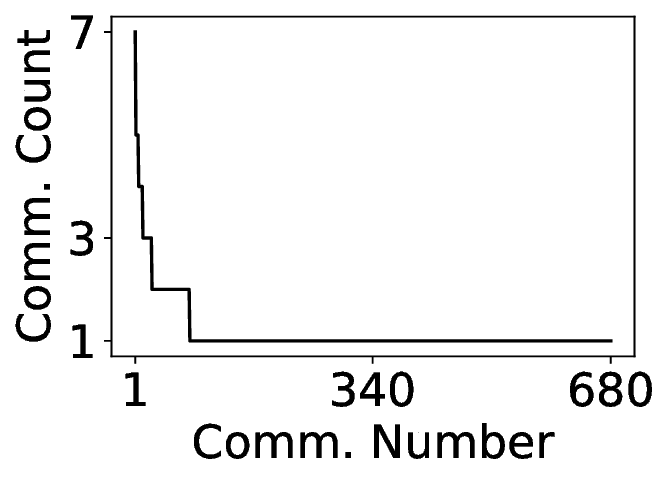} & 
\includegraphics[width=.23\textwidth]{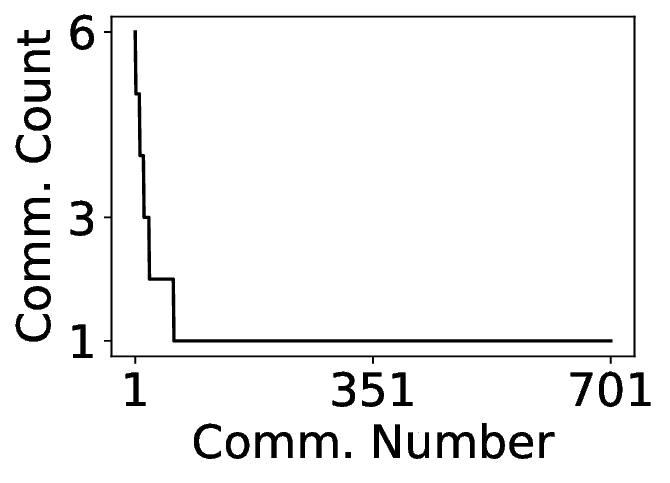} \\
\qquad(a) & \qquad(b) \\
\includegraphics[width=.23\textwidth]{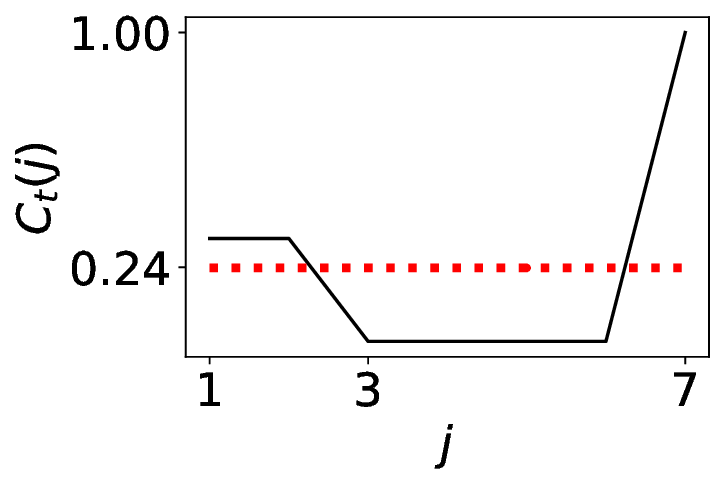} & 
\includegraphics[width=.23\textwidth]{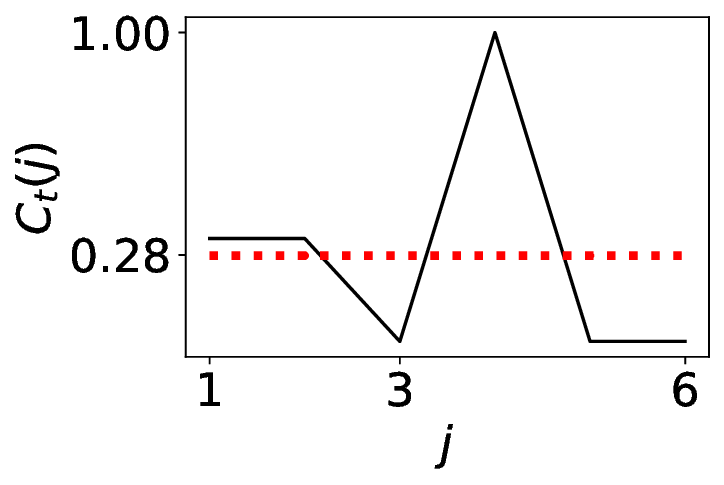} \\
\qquad(c) & \qquad(d) 
\end{tabular}
\caption{(a,b) The numbers of nodes in each connected component of ${\bf A}^{(\text{md})}$ and (c,d) the local clustering coefficients of the nodes in the LCC of ${\bf A}^{(\text{md})}$ of the BA and WS networks for oscillators with $\gamma = 10$. We show our results for a BA network in the left column and for a WS network in the right column. The large spread in oscillator frequencies for these strongly heterogeneous oscillators prevents the formation of large LCCs, and no community has more than a few oscillators in it. Nevertheless, the local clustering coefficients of the nodes in the LCC of ${\bf A}^{(\text{md})}$ are still larger than the baseline values in the original networks in Figure \ref{fig:fullgraphcluster}.  In (a) and (b), ``Comm." stands for ``Community".}
\label{fig:cluster_gam_ten}
\end{figure}

In Figure \ref{fig:order_param_gam_ten}, we see that the functional community of oscillators that consists of the LCC of ${\bf A}^{(\text{md})}$ is markedly closer to frequency locking than is the case for those oscillators are in the original network. This arises from the larger, less oscillatory values of $r_{p}$ and less erratic dynamics of $\cos(\psi_{p}(t))$ in Figures \ref{fig:order_param_gam_ten} (e)--(h) than in Figures \ref{fig:order_param_gam_ten} (a)--(d). Therefore, although few oscillators are in these communities when $\gamma = 10$ (with the vast majority of oscillators wandering incoherently), these lucky few oscillators are {much closer} to being locked together.

\begin{figure}[!h]
\centering 
\begin{tabular}{cc}
\includegraphics[width=.23\textwidth]{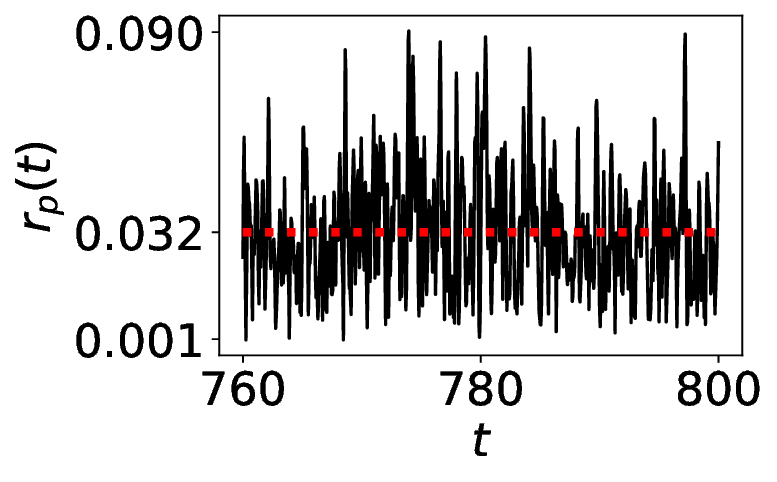} &  
\includegraphics[width=.23\textwidth]{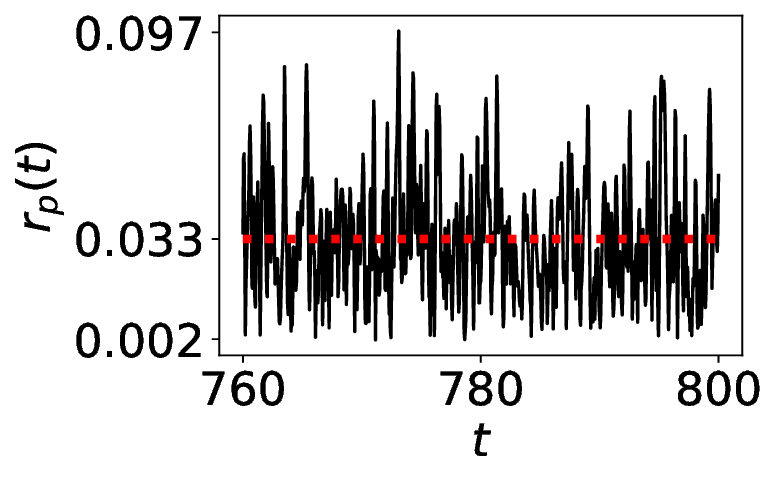}\\
\qquad(a) & \qquad(b) \\ 
\includegraphics[width=.23\textwidth]{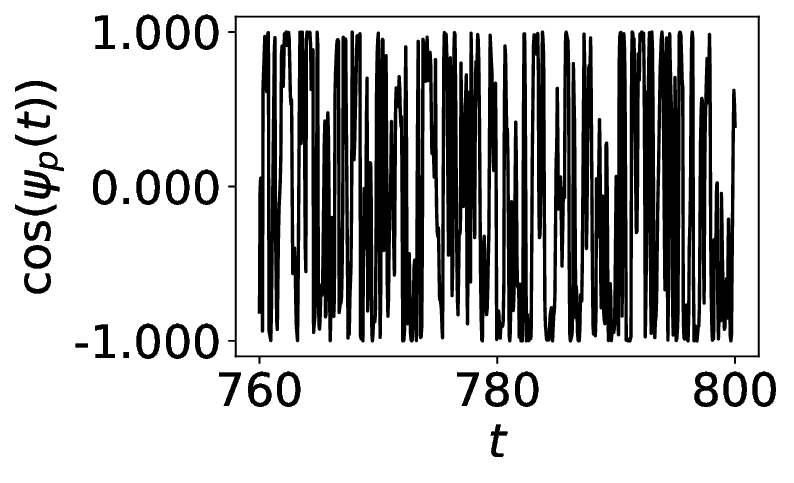} &  
\includegraphics[width=.23\textwidth]{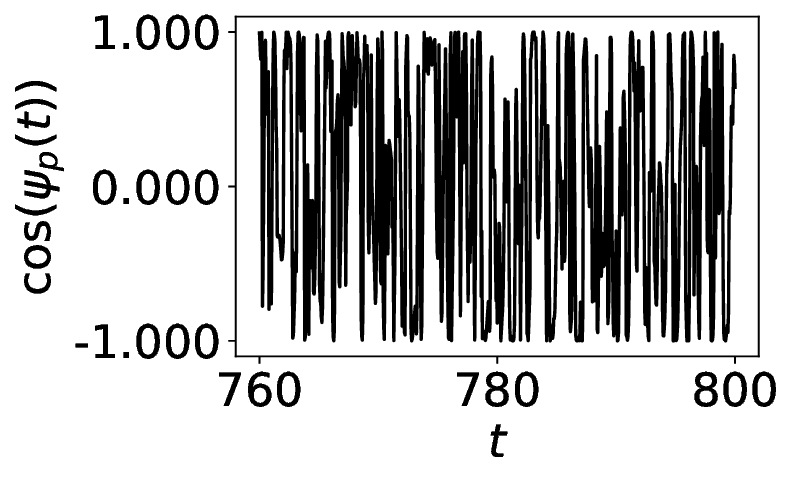}\\
\qquad(c) & \qquad(d) \\
\includegraphics[width=.23\textwidth]{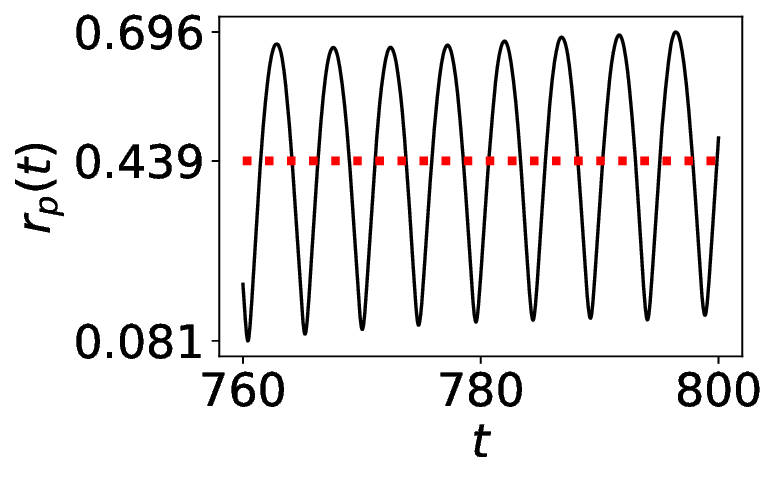}& 
\includegraphics[width=.23\textwidth]{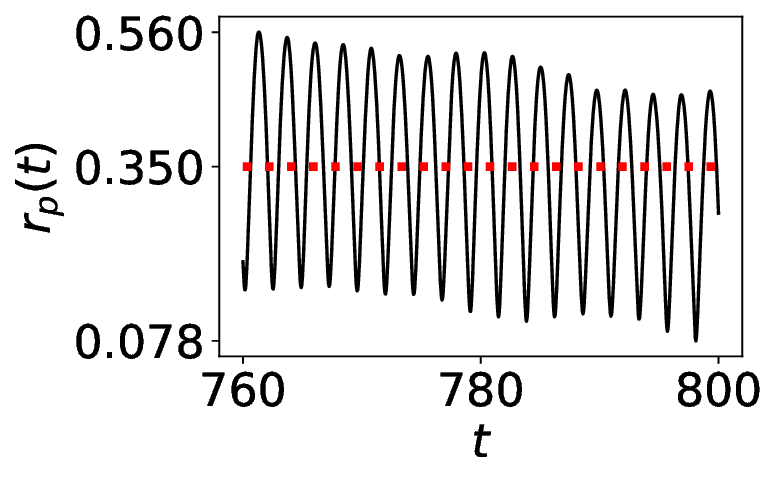}\\
\qquad(e) & \qquad(f) \\ 
\includegraphics[width=.23\textwidth]{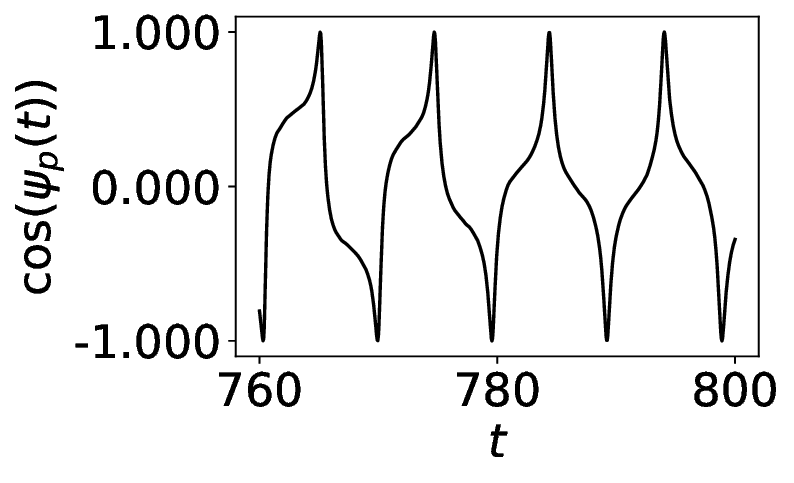}& 
\includegraphics[width=.23\textwidth]{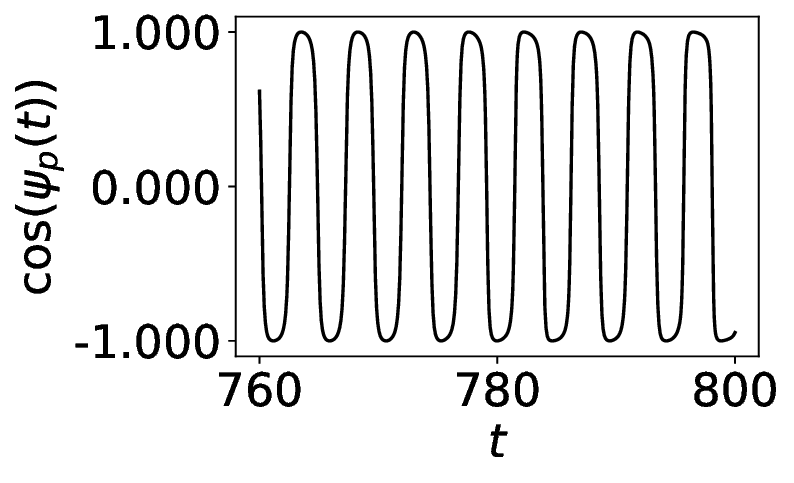}\\
\qquad(g) & \qquad(h) 
\end{tabular}
\caption{The (a,b,e,f) order-parameter magnitude $r_{p}(t)$ and (c,d,g,h) real part $\cos(\psi_{p}(t))$ of the order-parameter phase for the BA and WS networks when $\gamma=10$. Our results for the original two networks are in panels (a)--(d), and our results for the LCCs of ${\bf A}^{(\text{md})}$ of these networks are in panels (e)--(h). We show our results for a BA network in the left column and for a WS network in the right column. The horizontal line in (a,b,e,f) indicates the mean $\left\langle r_{p}(\cdot)\right\rangle$ of the order-parameter magnitude. The networks from both models have more frequency locking in the LCC of ${\bf A}^{(\text{md})}$ than in the original networks.}
\label{fig:order_param_gam_ten}
\end{figure}


\subsection{Community Dynamics}

Now that we have established that our method of generating functional communities of oscillators in a manner that is consistent with their dynamics, we compare how communities form when $\gamma = 0.1$ (i.e., the case of weakly heterogeneous oscillators) in the two RGMs. In Figure \ref{fig:comdynam}, we see that community formation occurs in starkly different ways in the two models, as the BA network has a far weaker tendency towards community coalescence than the WS network. We also see in the BA network that communities can cease to propagate in time, as there is often a sequence of coalescing communities that then stop propagating further. Therefore, our notion of functional communities allows us to examine their temporal evolution and thereby capture some details of the dynamics of the coupled oscillators.

\begin{figure}[!h]
\centering
\begin{tabular}{c}
\includegraphics[width=.45\textwidth]{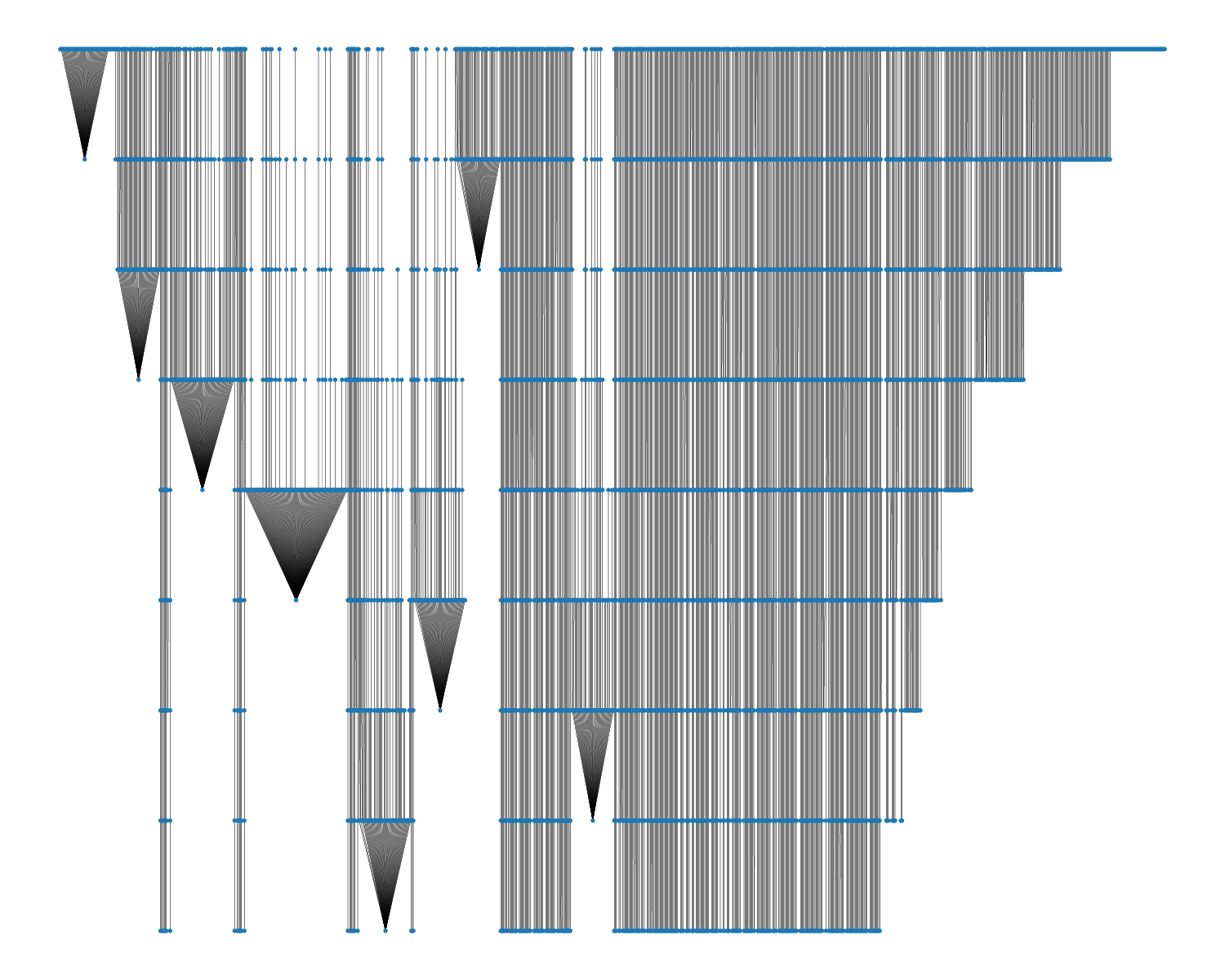}\\
(a) \\
\includegraphics[width=.45\textwidth]{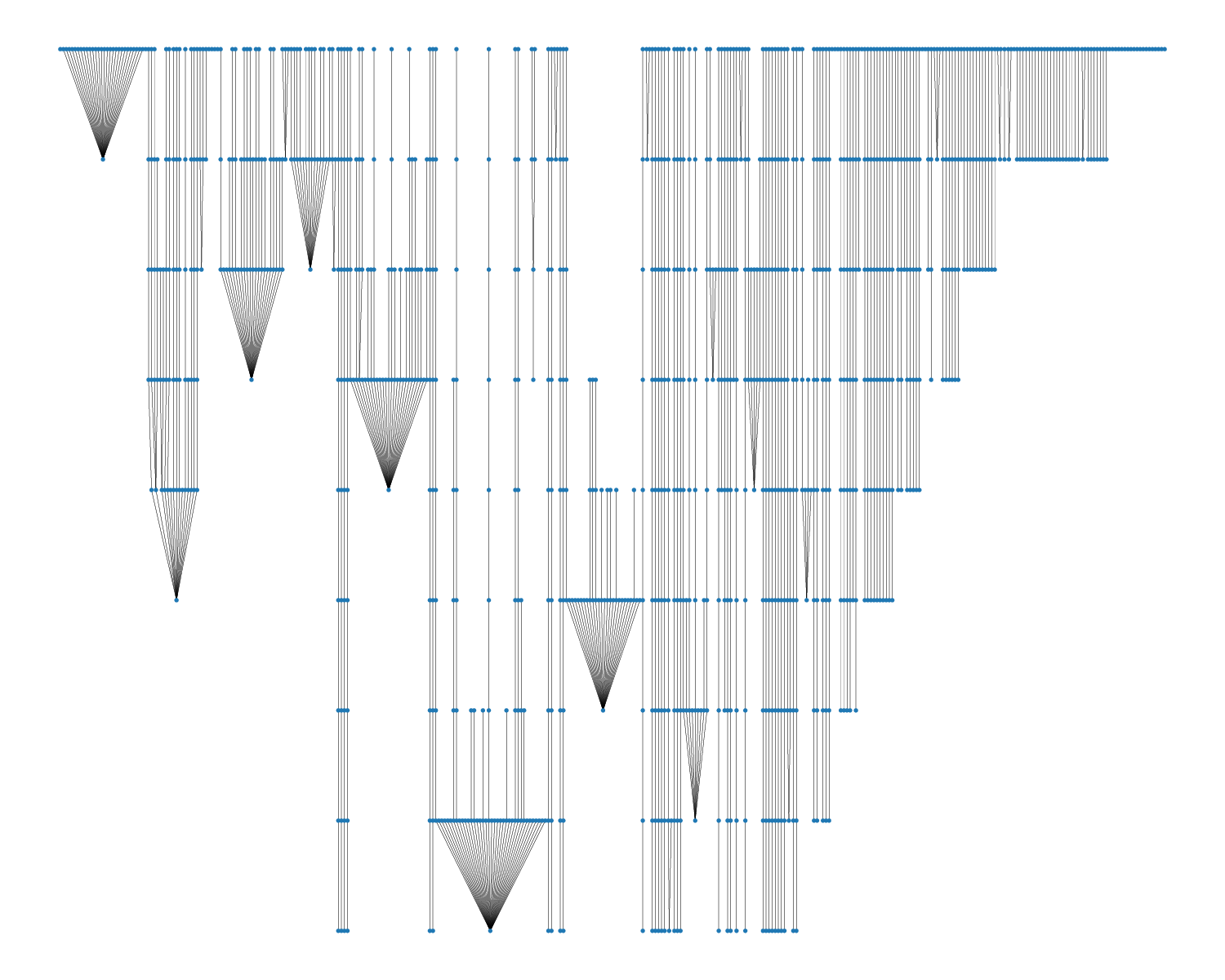}\\
(b) 
\end{tabular}
\caption{The community-relationship graphs $\mathcal{G}_{t_{i},t_{f}}$ for weakly heterogeneous oscillators (i.e., $\gamma = 0.1$) for (a) a BA network and (b) a WS network for the time interval $t_{i}=600 \leq t \leq t_{f}=800$ across overlapping intervals of $40$ time units (i.e., with $n_{w}=5$). Time runs from top to bottom. This plot illustrates how functional communities, which consist of oscillators with similar dynamics, evolve over time.}
\label{fig:comdynam}
\end{figure}


\section{Conclusions and Discussion} \label{sec5}

Using DMD, we developed a versatile approach for identifying functional communities of heterogeneous coupled oscillators on networks.  We examined coupled Kuramoto oscillators on networks that we constructed from random-graph models, and we found functional communities of oscillators based on how much they exhibit phase locking and frequency locking. These functional communities arise through clustering in graphs of community relationships, and the clustering coefficients in these graphs are larger than the clustering coefficients in the original networks. Unsurprisingly, in concert with associated synchronization properties, these functional communities are stronger for networks of weakly heterogeneous oscillators than they are for networks of strongly coupled oscillators. The community-relationship graphs, which take the form of forests, encode the interactions between the oscillators and provide a way to visualize their complicated dynamics over time. We observed coalescence of communities when oscillators exhibit phase locking or frequency locking. Additionally, from our community-relationship graphs, we observed that BA networks and WS networks yield functional communities with different community-coalescence properties.

Our results are promising, and it is important to test them in increasingly challenging scenarios. The Kuramoto model is very well-studied \cite{rodrigues}, so it useful to examine it as a starting point for developing approaches like ours.  However, it is desirable to challenge our approach with other models of coupled oscillators, other dynamical systems, and time-series output of natural observations and laboratory experiments. Additionally, we only examined coupled oscillators with uncorrelated natural frequencies, and it will be fascinating to apply our approach to examine dynamics in the presence of such correlations.

\section*{Acknowledgements}

We thank Arkady Pikovsky for helpful comments.


\bibliography{dmd_wavelet2}
\bibliographystyle{unsrt}

\end{document}